\documentclass[prb,showpacs,amsmath,amssymb]{revtex4}

\usepackage{graphicx}
\usepackage{dcolumn}

\renewcommand{\leq}{\leqslant}

\newcommand{\figurewidth}{0.48\textwidth}

\begin{document}

\title{Critical polymer--polymer phase separation in ternary solutions}

\author{Lei Guo}
\author{Erik Luijten}
\email[Corresponding author. E-mail: ]{luijten@uiuc.edu}
\affiliation{%
Department of Materials Science and Engineering,
University of Illinois at Urbana--Champaign,\\
1304 West Green Street, Urbana,
Illinois 61801, U.S.A.}

\date{\today}

\begin{abstract}
  We study polymer--polymer phase separation in a common good solvent by
  means of Monte Carlo simulations of the bond-fluctuation model. Below
  a critical, chain-length dependent concentration, no phase separation
  occurs. For higher concentrations, the critical demixing temperature
  scales nonlinearly with the total monomer concentration, with a power
  law relatively close to a renormalization-group prediction based on
  ``blob'' scaling arguments. We point out that earlier simulations and
  experiments have tested this power-law dependence at concentrations
  outside the validity regime of the scaling arguments. The critical
  amplitudes of the order parameter and the zero-angle scattering
  intensity also exhibit chain-length dependences that differ from the
  conventional predictions but are in excellent agreement with the
  renormalization-group results.  In addition, we characterize the
  variation of the average coil shape upon phase separation.
\end{abstract}  


\maketitle

\section{Introduction}

Understanding the incompatibility of chemically different polymer
species is a topic of great practical importance that consequently has
received widespread attention over the last decades. Although a very
large number of experiments has been performed---including determination
of the coexistence curve\cite{esker76a}, the Flory--Huggins interaction
parameter\cite{fukuda84,fukuda86,kent92}, and of critical
exponents\cite{nose95,schwahn02}---several basic theoretical
predictions~\cite{scott49,flory53,degennes79,joanny84,schaefer85,broseta87}
have proven difficult to verify, as they rely on specific assumptions,
such as monodisperse and symmetric conditions, in which the solvent
interacts equally with both polymer species. While, on the other hand,
these conditions can easily be realized in computer simulations, those
mostly have focused on polymer demixing in binary blends, i.e., at
relatively high concentrations (cf.\ 
Refs.~\onlinecite{deutsch92,deutsch93b}). Sariban and
Binder~\cite{sariban87,sariban88b,sariban94} have performed pioneering
computational work on polymer--polymer demixing under more dilute
conditions, using a simple cubic lattice model. The fact that little
additional computational work appears to have been done since then has
motivated us to further explore polymer demixing in the presence of a
solvent.

Flory--Huggins~(FH) theory~\cite{scott49,flory53} predicts that two
incompatible polymer species in a nonselective good solvent will phase
separate at arbitrary dilution, provided the temperature of the system
is sufficiently low. Under the appropriate conditions, the phase
transition will be continuous and the corresponding critical
temperature~$T_{\rm c}$ is predicted to increase linearly with the total
monomer concentration~$\phi$ over the entire concentration range.
However, this mean-field approach assumes homogeneous monomer densities
for both species and ignores the chain connectivity. Below the
dilute--semidilute threshold~$\phi^*$, this connectivity causes
individual chains to be separate, swollen coils that do not
interpenetrate. Thus, de Gennes~\cite{degennes79} has claimed that
\emph{no phase separation} occurs for concentrations $\phi < \phi^*$.
Furthermore, in the semidilute regime, $\phi > \phi^*$, each chain can
be viewed as a succession of ``blobs,'' where each blob contains several
monomers. In a good solvent, a blob does not contain monomers of other
chains due to the excluded-volume interactions. This reduces the number
of contacts between monomers of different species and consequently
lowers the critical temperature compared to the FH prediction.
Specifically, the critical temperature no longer varies linearly with
concentration, but decreases superlinearly with decreasing
concentration.\cite{degennes78,degennes79} Employing
renormalization-group~(RG) techniques, Joanny and
co-workers\cite{joanny84} observed that demixing under good solvent
conditions is driven by corrections to scaling. Sch\"afer and
Kappeler~\cite{schaefer85} calculated the corresponding spinodal by
means of an RG approach and found a nonlinear dependence of the critical
temperature on concentration that differs from the prediction of
Ref.~\onlinecite{degennes78}.  Broseta \emph{et al.}~\cite{broseta87}
subsequently extended the RG treatment to predict the chain-length
dependences of critical amplitudes of the order parameter, zero-angle
scattering intensity and the correlation length. These predictions have
partially been tested in Ref.~\onlinecite{sariban94}, where the
concentration dependence of the critical temperature in the semidilute
regime and the chain-length dependence of the critical amplitude of the
order parameter have been investigated.

In this paper, we extend the work of Ref.~\onlinecite{sariban94} by
means of Monte Carlo simulations in which we employ the bond-fluctuation
model~\cite{carmesin88,deutsch91} rather than a simple cubic lattice
model. This model is capable of a more realistic representation of
mixtures of flexible chains, as it permits variation of the length of
chain segments and a much smoother variation of the angle between
successive segments.  We investigate phase separation as a function of
total monomer concentration and test the various theoretical predictions
for the critical properties, where we also address an alternative
scenario leading to a nonlinear variation of the transition temperature
with concentration.  In addition, our simulations explicitly cover the
transition from the dilute to the semidilute regime in ternary
solutions, which requires (computationally demanding) calculations in
the grand-canonical ensemble.  Following our earlier work~\cite{guo03a},
we also investigate the effect of this transition on shape properties of
individual polymer coils.  The shape variation of polymers upon phase
separation is of interest because of its influence on the physical
properties of ternary solutions. As was first recognized by
Kuhn~\cite{kuhn34}, the shape of a flexible polymer is ellipsoidal
rather than spherical. This may affect, e.g., the flow properties of
polymeric fluids~\cite{abernathy80} and the polymer-induced depletion
potential in colloid--polymer mixtures.\cite{kamien99} Furthermore,
Murat and Kremer~\cite{murat98} proposed a coarse-grained model for the
study of phase separation of polymer blends, in which each coil is
replaced by a ``soft'' ellipsoidal particle.  Since individual segments
are no longer modeled explicitly, this may permit the observation of the
demixing process at longer times scales.  However, it also requires
accurate knowledge of the actual ellipsoidal shape of the individual
chains.  For polymer chains in the dilute limit, corresponding studies
have indeed been performed. \v{S}olc and
Stockmayer~\cite{stockmayer71,solc71} first introduced the
radius-of-gyration tensor. The three eigenvalues of this gyration tensor
are the squared principal components of the radius of gyration. From
Monte Carlo simulations of an ideal polymer chain on a cubic lattice,
they found that these three components were very different, implying an
asymmetric polymer shape and confirming Kuhn's original observation.
The \emph{asphericity}~$A$ was introduced to characterize the coil
shape.\cite{theodorou85,nelson86} It takes values between $0$ (sphere)
and $1$ (rod) and was calculated for ideal and self-avoiding
chains.\cite{rudnick86,nelson86} Since this asphericity does not
distinguish prolate-ellipsoid and oblate-ellipsoid shapes, another
parameter~$S$ was introduced~\cite{nelson86} and calculated analytically
for ideal and self-avoiding chains. However, in order to simplify the
calculations for $A$ and~$S$ (both of which are defined below),
expectation values of ratios were replaced by ratios of averages.  This
approximation was found to overestimate the asphericity of polymer
chains.\cite{diehl89,kremer92} Numerical
calculations~\cite{bishop86,goldbart91,kremer92} have supported the
analytical calculations.  In recent years, it has actually become
possible to observe the shape asphericity
experimentally.\cite{wirtz00,maier01} These experiments, as well as the
vast majority of the theoretical work, focus on the shape of a single
polymer chain in a highly dilute solution.  Studies of the influence of
concentration are rare and have essentially shown that the asphericity
of athermal chains diminishes only very gradually as a function of
increasing concentration.\cite{bishop80,zifferer97} In a poor solvent,
the reverse effect occurs, due to the coil--globule transition taking
place in the polymer-lean phase.\cite{szleifer90} The dependence of
polymer shape on both concentration and solvent quality has motivated
our earlier study of the shape change of polymers upon phase separation
in ternary solutions.\cite{guo03a} It was found that phase separation
strongly influences the shape of the minority component in a given
phase. Here, we expand this work by investigating the temperature
dependence of both the asphericity~$A$ and the prolate--oblate
parameter~$S$ at fixed total concentration.

\section{Theoretical background}
\label{sec:theory}

Our study consists of three main aspects: (i)~The scaling of the
critical temperature $T_{\rm c}$ with total monomer concentration $\phi$
and degree of polymerization $N$; (ii)~the chain-length dependence of
critical amplitudes; (iii)~the shape variation of polymers upon phase
separation. In this section, we provide the necessary theoretical
background and review the corresponding predictions.

\subsection{Scaling of the critical demixing temperature with
concentration and degree of polymerization}
\label{sec:theory_tc}

We consider a monodisperse solution containing polymers of two different
species, denoted A and~B, with degree of polymerization $N_{\rm
A}=N_{\rm B}=N$. In addition, the solvent is of the same quality for
both species, and their chemical potentials are equal. Thus, a so-called
\emph{symmetric} ternary system is realized, in which both species
behave identically. The only distinction between monomers of type~A and
monomers of type~B is their mutual repulsion of
strength~$\varepsilon_{\rm AB}>0$.  Identical monomers experience an
attraction $\varepsilon_{\rm AA}=\varepsilon_{\rm BB}<0$.  Accordingly,
phase separation occurs at sufficiently low temperatures.  Owing to the
symmetric properties of both species, this transition is continuous if
the chemical potential difference vanishes.  FH theory, assuming
complete screening of the excluded-volume interactions, predicts a
linear relation between the corresponding critical temperature $T_{\rm
c}$ and the monomer concentration~$\phi$ (cf.\ 
Ref.~\onlinecite{flory53}),
\begin{equation}
\frac{\Delta\varepsilon}{k_{\rm B} T_{\rm c}} \propto
\frac{1}{N}\phi^{-1} \;,
\label{eq:FH_Tc}
\end{equation}
where $\Delta\varepsilon=(2\varepsilon_{\rm AB}-\varepsilon_{\rm
AA}-\varepsilon_{\rm BB})/2$.  Equation~(\ref{eq:FH_Tc}) implies that
polymer--polymer (PP) phase separation occurs for arbitrarily
low~$\phi$, provided that the temperature is sufficiently low. However,
in the dilute regime ($\phi<\phi^*$), the excluded-volume interaction is
not screened at all and polymer coils essentially do not interact.
Accordingly, de Gennes~\cite{degennes78,degennes79} argued that no PP
phase separation can occur for concentrations below $\phi^*$. At the
overlap threshold~$\phi^*$ phase separation can be induced, provided
that the segregation factor is sufficiently strong (or the temperature
sufficiently low).  In the semidilute regime ($\phi>\phi^*$), the
tendency to phase separate is suppressed as well, and demixing will take
place at a lower temperature than predicted by FH theory, due to the
partial screening of the excluded-volume interaction.  According to blob
scaling arguments~\cite{degennes79}, in this regime each chain can be
viewed as a succession of blobs. Each blob only contains monomers of a
single species, so that the number of A--B contacts is reduced and phase
separation is correspondingly suppressed.  Indeed, de Gennes has
predicted the critical temperature to scale
as\cite{degennes78,degennes79}
\begin{equation}
\label{eq:crittemp}
\frac{\Delta\varepsilon}{k_{\rm B} T_{\rm c}} \propto
\frac{1}{N}\phi^{-\frac{1}{3\nu-1}}
\simeq \frac{1}{N}\phi^{-1.31} \quad (\phi>\phi^*) \;.
\end{equation}
Alternatively, phase separation can be induced by varying the
concentration at a fixed temperature~$T$. At low temperatures, it
follows from the previous argument that the total monomer concentration
must be increased to~$\phi^*$.  However, if the segregation factor is
weak (i.e., at high temperatures), the homogeneous phase persists into
the semidilute regime and separation only occurs at a critical
concentration
\begin{equation}
\label{eq:critconc}
  \phi_{\rm c} \propto (T/N)^{3\nu-1} \simeq (T/N)^{0.764} \quad
  (\phi>\phi^*) \;,
\end{equation}
where $\nu = 0.588$ is the scaling exponent for the end-to-end
distance~$R$ (or the radius of gyration~$R_g$) as a function of $N$ in
dilute solution. We observe that $\phi_{\rm c}$ exhibits the same
chain-length dependence as the overlap threshold~$\phi^*$.  By contrast,
an RG approach~\cite{joanny84} indicated that for the fixed point
corresponding to phase separation under good solvent conditions, the
chemical mismatch between A and B is an \emph{irrelevant parameter}.
Indeed, phase separation is predicted to be driven by the
\emph{corrections to scaling}.  Consequently, the scaling predictions
Eqs.\ (\ref{eq:crittemp}) and~(\ref{eq:critconc}) are modified, as shown
by an RG calculation of the spinodal.~\cite{schaefer85} Specifically,
it was found that for demixing in the semidilute regime, at a
temperature~$T$, the critical concentration scales as,
\begin{equation}
\label{eq:critconc-rg}
  \phi_{\rm c} \propto (T/N)^{(3\nu-1)/(1+x)} \simeq (T/N)^{0.624} \;.
\end{equation}
From the analogy between a polymer solution and the $n$-vector model in
the limit $n \to 0$, the exponent $x$ can be related to the negative of
the crossover exponent of the isotropic fixed point of the $n$-vector
model with cubic anisotropy.\cite{joanny84,broseta87} This crossover
exponent has been calculated in a third-order
$\varepsilon$-expansion~\cite{ketley73}, and from a Pad\'e--Borel
resummation the value $x=0.225~(5)$ was obtained.\cite{broseta87} The
most notable difference between Eqs.\ (\ref{eq:critconc})
and~(\ref{eq:critconc-rg}) is that in the RG result the critical
concentration (at fixed temperature) decreases with chain length at a
slower rate than the overlap threshold $\phi^* \sim 1/N^{3\nu-1}$, i.e.,
for sufficiently long chains (and weak segregation factor) phase
separation sets in at higher concentrations than predicted by
Eq.~(\ref{eq:critconc}).  For experiments at fixed concentration [cf.\ 
Eqs.\ (\ref{eq:FH_Tc}) and~(\ref{eq:crittemp})], the critical
temperature is now given by
\begin{equation}
\frac{\Delta\varepsilon}{k_{\rm B}T_{\rm c}} \propto
\frac{1}{N}\phi^{-\frac{1+x}{3\nu-1}} \simeq \frac{1}{N}\phi^{-1.60} \quad
(\phi>\phi^*) \;.
\label{eq:RG_Tc}
\end{equation}

Interestingly, Olvera de la Cruz\cite{olvera89} generalized the work of
Ref.~\onlinecite{broseta87} to microphase separation in diblock
copolymer solutions and predicted that the order--disorder transition
temperature $T_{\rm ODT}$ exhibits the same concentration dependence as
described in Eq.~(\ref{eq:RG_Tc}). Lodge \emph{et al.}\cite{lodge95b}
found very good agreement with this prediction for solutions of
poly(styrene-\emph{b}-isoprene) (PS-PI) diblock copolymers, whereas the
transition temperatures for
poly(ethylenepropylene-\emph{b}-ethylethylene) solutions followed a
\emph{different} power law. While the results for PS-PI solutions can be
viewed as a confirmation of the RG scenario, the agreement between
theory and experiment may well be fortuitous, as most data were taken in
the concentrated regime where the blob scaling approach is invalid.
Guenza and Schweizer\cite{guenza97} subsequently studied the
order--disorder transition by means of PRISM theory and observed that
local concentration fluctuation effects also imply a nonlinear variation
of~$T_{\rm ODT}$, resulting from a concentration-dependent local
correlation hole affecting the $\chi$ parameter,
\begin{equation}
\frac{1}{T_{\rm ODT}} \propto \frac{1}{N}\phi^{-\frac{4\nu-1}{3\nu-1}}
\simeq \frac{1}{N}\phi^{-1.77} \;.
\label{eq:PRISM_Tc}
\end{equation}
Thus, while the integral equation theory cannot quantitatively capture
the critical fluctuations that lead to the RG result, it predicts a
numerically similar power law that results from a very different
mechanism. Unlike the blob scaling arguments, this mechanism explicitly
applies to the concentrated solution regime.

\subsection{Scaling of critical amplitudes with degree of polymerization}

Critical properties only exhibit nonclassical scaling behavior within a
certain region around the critical point, determined by the Ginzburg
criterion.\cite{ginzburg} For larger deviations from $T_{\rm c}$, i.e.,
when $|t| \gg G$, where $t$ is defined as $(T-T_{\rm c})/T_{\rm c}$ and
$G$ denotes the Ginzburg number, classical or mean-field-like critical
exponents are observed (cf.\ Refs.~\onlinecite{chi3d,mr3d}). For polymer
mixtures, the Ginzburg number decreases as $1/N$, so that Ising-like
critical exponents are only observed in a very narrow temperature region
around the critical point.\cite{degennes77} According to Broseta
\emph{et al.}~\cite{broseta87}, the modified chain-length dependence of
the critical temperature and critical concentration described in
Sec~\ref{sec:theory_tc} also affects the Ginzburg criterion, so that the
number of blobs per chain~$\tilde{N} \propto (\phi_{\rm
c}/\phi^*)^{1/(3\nu-1)} \sim N^{x/(1+x)}$ replaces $N$. Thus,
nonclassical behavior is observed within the considerably larger region
$|t| N^{x/(1+x)} \ll 1$. In addition, the chain-length dependence of all
critical amplitudes is modified. In this work, we test these predictions
for the order parameter as well as for the zero-angle scattering
intensity, for which---to our knowledge---it has not been verified
before.  In our \mbox{(semi-)}grand-canonical simulations (see
Sec.~\ref{sec:model}), the order parameter~$m$ is defined as
\begin{equation}
m=\frac{n_{\rm A}-n_{\rm B}}{n_{\rm A}+n_{\rm B}} \;,
\end{equation}
where $n_{\rm A}$, $n_{\rm B}$ are the numbers of A and B polymers,
respectively.  This corresponds to the concentration difference (for
either species~A or species~B) between the A-rich and the B-rich phases.
In simulations of finite systems, the ensemble average $\langle
m\rangle$ vanishes~\cite{binder01}, which is resolved by employing the
absolute value of the order parameter, $\langle |m| \rangle$.  For the
sake of consistency, we thus phrase the following theoretical
expressions in terms of $\langle |m|\rangle$. Composition fluctuations
are probed via the zero-angle scattering intensity (osmotic
compressibility)~\cite{hill56}
\begin{equation}
S_{\rm coll}(0) = nN\phi (\langle |m|^2\rangle -\langle |m|\rangle ^2)
  \propto \phi^2 V (\langle |m|^2\rangle -\langle |m|\rangle ^2) \;,
\label{eq:scoll}
\end{equation}
where $n\equiv n_{\rm A}+n_{\rm B}$ is the total number of polymers.
Near criticality, both $\langle |m|\rangle$ and $S_{\rm coll}(0)$
exhibit a power-law dependence on the reduced temperature~$t$,
\begin{eqnarray}
\label{eq:mN_prediction}
\langle |m|\rangle &=& \hat{B}(N) |t|^{\beta} \;, \\
\label{eq:SN_prediction}
S_{\rm coll}(0)    &=& \hat{\Gamma}(N) |t|^{-\gamma} \;,
\end{eqnarray}
where $\beta$ and $\gamma$ denote critical exponents and $\hat{B}(N)$
and $\hat{\Gamma}(N)$ are $N$-dependent amplitudes. Within the
asymptotic scaling region, $\beta$ and $\gamma$ assume their Ising
values, $\beta \simeq 0.327$ and $\gamma \simeq 1.237$.\cite{ising3d}
The chain-length dependence of the amplitudes follows directly from the
observation that $t/G = tN$ constitutes the proper scaling variable and
that $\hat{B}$ and $\hat{\Gamma}$ must be independent of~$N$ in the
mean-field limit. Thus,
\begin{alignat}{3}
\label{eq:mN}
\hat{B}(N) &\propto N^{\beta-1/2}     &&\simeq N^{-0.173} \;,\\
\label{eq:SN}
\hat{\Gamma}(N) &\propto N \cdot N^{1-\gamma} &&\simeq N^{0.763} \;,
\end{alignat}
where $\hat{\Gamma}(N)$ involves an additional factor~$N$ because of the
prefactor in Eq.~(\ref{eq:scoll}).  The revised scaling of $\hat{B}(N)$
and $\hat{\Gamma}(N)$ with $N$ yields~\cite{broseta87,sariban94}
\begin{alignat}{4}
\label{eq:mN_RG}
\hat{B}(N)      &\propto \tilde{N}^{\beta-1/2}
                &&= N^{x(\beta-1/2)/(1+x)}  &&\simeq N^{-0.0318} \;,\\
\label{eq:SN_RG}
\hat{\Gamma}(N) &\propto N \cdot \tilde{N}^{1-\gamma}
                &&= N^{1+x(1-\gamma)/(1+x)} &&\simeq N^{0.956} \;.
\end{alignat}
Equation~(\ref{eq:mN_RG}) was verified in Ref.~\onlinecite{sariban94}.
It is one of the goals of the present work to test the prediction
Eq.~(\ref{eq:SN_RG}) for the compressibility and to reproduce the
scaling for the order parameter in the context of the bond fluctuation
model. A final noteworthy point concerns the so-called Fisher
renormalization of critical exponents\cite{fisher68}, which is expected
to occur for experiments and simulations at fixed concentration. It has
been predicted\cite{broseta87} that for polymer demixing this
renormalization occurs within an exceedingly narrow range around the
critical temperature, whereas regular Ising-type exponents are predicted
to occur outside this range (but within the nonclassical regime
predicted by the Ginzburg criterion).

\subsection{Shape variation of polymer coils upon phase separation}

The ellipsoidal shape of a polymer coil is characterized by the
eigenvalues $\lambda_1 \leq \lambda_2 \leq \lambda_3$ of the
radius-of-gyration tensor $\mathbf{Q}$, which is defined
as~\cite{solc71,rudnick86}
\begin{equation}
Q_{\alpha\beta} = \frac{1}{2N^2}
\sum_{i,j=1}^{N}[r_{i,\alpha}-r_{j,\alpha}][r_{i,\beta}-r_{j,\beta}] \;,
\end{equation}
where $\mathbf{r}_{i}$ represents the position of the $i$th monomer
along the chain and $\alpha,\beta=1,2,3$ denote cartesian components.
An important measure is the \emph{asphericity}
$A$~\cite{nelson86,gaspari87,diehl89,kremer92}:
\begin{equation}
A = \frac{1}{2} \left\langle
\frac{(\lambda_1-\lambda_2)^2 + (\lambda_2-\lambda_3)^2 +
(\lambda_3-\lambda_1)^2}{(\lambda_1+\lambda_2+\lambda_3)^2} \right\rangle \;,
\label{eq:A}
\end{equation}
which takes values between 0 (sphere) and 1 (rod).  In dilute solution
it approaches a universal value as $N\to\infty$, estimated as $0.415$
from first-order $\varepsilon$-expansions and as $0.431$ from
simulations.\cite{bishop88,goldbart91,kremer92} In the melt limit,
where the chains behave ideally, this value is anticipated to decrease
to the (exactly known) RW value $0.39427\ldots$.\cite{diehl89} Another
measure of the polymer shape is $S$,
\begin{equation}
S=\left\langle \frac{(\lambda_1-\overline{\lambda})
  (\lambda_2-\overline{\lambda})
  (\lambda_3-\overline{\lambda})}{{\overline{\lambda}}^3} \right\rangle
  \;,
\label{eq:S}
\end{equation}
where $\overline{\lambda}$ is the average value of $\lambda_1$,
$\lambda_2$ and $\lambda_3$. $S$ distinguishes oblate-shaped polymers
($-\smash{\frac{1}{4}}<S<0$) from prolate-shaped chains ($0<S<2$). In
the dilute limit it approaches a universal value $0.541$, as estimated
from simulations.\cite{goldbart91,kremer92} Upon approach of the melt
limit it decreases to a value that is estimated as $0.475$ from
numerical integration~\cite{kremer92} and $0.478$ from
simulations.\cite{kremer92}

\section{Simulation model and details}
\label{sec:model}

The bond fluctuation model (BFM) has been introduced for the Monte Carlo
simulation of polymer systems in Refs.\ \onlinecite{carmesin88}
and~\onlinecite{deutsch91}. In this lattice model, a chain consists of
connected units, each of which occupies a cubic lattice cell.  Every
unit represents a Kuhn segment corresponding to 3 to 5 real
monomers.\cite{tries97} The units cannot overlap and are connected by
bond vectors with lengths between 2 and $\sqrt{10}$ lattice constants;
restricting the vectors to this set prevents the crossing of bonds.
Important differences between the BFM and a self-avoiding random-walk
(Verdier--Stockmayer~\cite{verdier62}) model are the flexibility of the
segment length (the distance between two connected units) and the
resulting increase in the number of possible bond angles (i.e., the
angle between two adjacent segments).  Monomer interactions are
implemented by means of a square-well potential with a range of
$\sqrt{6}$ lattice constants, which covers 54 out of the 108 possible
positions for neighboring segments. The attractive coupling strength
between segments of the same species is set to $\varepsilon_{\rm
AA}=\varepsilon_{\rm BB}=-1/k_{\rm B}T$.  Unlike species repel each
other with a strength $\varepsilon_{\rm AB}=\delta/k_{\rm B}T$, where
$\delta > 0$ is a variable parameter.  Throughout this paper, units are
chosen such that $k_{\rm B}=1$ and temperatures correspond to the
inverse coupling constant $\delta/\varepsilon_{\rm AB} =
-1/\varepsilon_{\rm AA}$.  The solvent is represented by empty lattice
sites, and the polymer--solvent interaction~$\varepsilon_{\rm AS}$ and
the solvent--solvent interaction~$\varepsilon_{\rm SS}$ both vanish.

The ternary system described here can undergo both polymer--polymer~(PP)
and polymer--solvent~(PS) demixing. Although the former typically occurs
at a higher temperature than the latter, it follows from the predictions
of Sec.~\ref{sec:theory} that the PP demixing temperature decreases with
decreasing concentration. Indeed, we found that for simulations using
the above-mentioned interaction parameters PP demixing is preempted by
PS separation near the dilute--semidilute threshold~$\phi^*$.  In order
to be able to determine the concentration dependence of the PP demixing
temperature, we therefore explicitly suppress PS phase separation in a
\emph{subset} of our calculations. The data presented in Sections
\ref{sec:phi_star} and~\ref{sec:shape} (with the exception of
Fig.~\ref{fig:N20_A3_thermal}) pertain to $\varepsilon_{\rm
AA}=\varepsilon_{\rm BB}=0$. This eliminates all attractive interactions
and, apart from the excluded-volume interactions, only leaves the A--B
repulsion nonzero.

All simulations are performed for symmetric, monodisperse systems
($N_{\rm A}=N_{\rm B}=N$) on simple cubic lattices with linear
dimension~$L$ and periodic boundary conditions. In order to permit a
finite-size scaling analysis for the determination of critical
properties, four different values for $L$ are studied.  Most simulations
are performed in the semi-grand-canonical ensemble~\cite{sariban87}, in
which the \emph{total} monomer concentration $\phi = \phi_{\rm A} +
\phi_{\rm B}$ is kept constant and only the identity of chains is
changed, governed by their interactions in a given configuration as well
as their chemical potential difference.  The symmetry of the system
makes such Monte Carlo moves possible and also guarantees that the
critical point will occur for identical chemical potentials. The A--B
changes are supplemented by local (translational) segment moves and
reptation-like moves. For strong repulsions, polymer demixing occurs
upon variation of concentration rather than temperature (this is the
predicted demixing near $\phi = \phi^*$, see Sec.~\ref{sec:theory_tc}).
Therefore, these simulations are carried out in the full grand-canonical
ensemble, in which the \emph{total} monomer concentration fluctuates as
well.  These calculations, which are particularly computationally
intensive, are performed using a variant of the recoil-growth
scheme.\cite{consta99,luijten03a} The chemical potential of both species
is varied, but the chemical potential difference is fixed at zero in
order to maintain critical demixing.

In the semi-grand-canonical simulations, all properties are sampled
every 50 or 200 sweeps (depending on $\phi$ and~$T$), where a sweep
corresponds to a sequence of, on average, three reptation moves per
chain, one local move per monomer and a semi-grand-canonical move for
one quarter of all chains.  After equilibration, which comprises
$4\,000$ samples, $100\,000$ samples are obtained for each state point.
For the grand-canonical simulations, properties are sampled every sweep,
which corresponds to 200 attempts to insert or delete a polymer chain
followed by a semi-grand-canonical move for half of all the chains. In
this case, equilibration corresponds to 500 sweeps and each production
run to $20\,000$ to $60\,000$ sweeps.  The data are analyzed by means of
multiple-histogram reweighting.\cite{multiple-hist} For the
semi-grand-canonical data only temperature reweighting is performed,
whereas the grand-canonical data also permit reweighting with respect to
the total monomer concentration (cf.\ Figs.\ \ref{fig:chi_L}a
and~\ref{fig:Q_phi}).

\section{Simulation results}
\label{sec:results}

\subsection{Scaling of the critical demixing temperature in
semidilute solutions}
\label{sec:Tc}

\begin{table}
\caption{Summary of linear system sizes~$L$ studied for different chain
lengths~$N$ and concentrations~$\phi$. Each system contains 
$n=\phi L^3/(8N)$ chains.}
\label{tab:systems}
\begin{ruledtabular}
\begin{tabular}{cccccccc}
\mbox{} & $\phi=0.12$ & $\phi=0.16$ & $\phi=0.20$ & $\phi=0.24$ &
$\phi=0.28$ & $\phi=0.32$ & $\phi=0.40$ 
\\ \hline 
$N=10$ & & & 20, 40, 60, 80 & 30, 40, 50, 60 & 20, 40, 60, 80 &
30, 40, 50, 60 & 20, 30, 40, 50
\\ 
$N=20$ & 40, 60, 80, 100 & 40, 60, 80, 100 & 40, 60, 80, 100 & 
40, 60, 80, 100 & 40, 60, 80, 100 & 40, 60, 80, 100
\\
$N=40$ & 60, 80, 100, 120 & 60, 80, 100, 120 & 40, 60, 80, 100 &
40, 60, 80, 100
\end{tabular}
\end{ruledtabular}
\end{table}

\begin{figure}
\begin{center}
\includegraphics[width=\figurewidth]{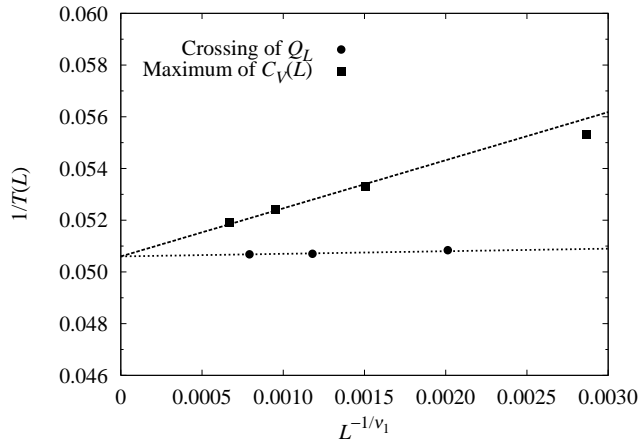}
\caption{\label{fig:Tc_L}Finite-size scaling behavior of the
characteristic temperatures~$T(L)$ determined from crossing points of
the fourth-order amplitude ratio~$Q_L$ and from maxima of the specific
heat~$C_{V}$. Data apply to a ternary polymer solution with chain
length~$N=20$, total monomer concentration~$\phi=0.32$ and repulsion
parameter~$\delta=1$. Both cases extrapolate to virtually identical
estimates for the critical demixing temperature~$T_{\rm c}$.}
\end{center}
\end{figure}

In order to determine the critical properties, we first locate the
critical temperature~$T_{\rm c}$ for the polymer--polymer phase
separation, as a function of concentration~$\phi$ and chain length~$N$.
For $N=10$, $20$ and~$40$ we have simulated cubic cells containing up to
$1792$, $2000$ and $750$ chains, respectively. Table~\ref{tab:systems}
lists the system sizes and concentrations employed.  We locate $T_{\rm
c}$ by means of the finite-size scaling properties of the fourth-order
amplitude ratio~$Q_L$~\cite{binder81},
\begin{equation}
Q_L=\langle m^2\rangle ^2/\langle m^4\rangle
\label{eq:Q_L}
\end{equation}
and the specific heat $C_V(L)$,
\begin{equation}
C_V(L) = (\langle E^2\rangle-\langle E\rangle ^2)/k_{\rm B}T^2 \;.
\label{eq:C_L}
\end{equation}
For each pair of system sizes $L_1$ and~$L_2$, the curves for $Q_L$ as a
function of temperature exhibit a crossing point that defines a
characteristic temperature~$T_Q(L)$, where we choose $L=(L_1+L_2)/2$.
Likewise, $C_V(L)$ exhibits a maximum at a temperature $T_C(L)$.  Both
characteristic temperatures approach the critical temperature in the
thermodynamic limit. Figure~\ref{fig:Tc_L} illustrates this for a
typical system with $N=20$, $\phi=0.32$ and $\delta=1$. In this figure,
we also exploit the theoretically predicted leading finite-size scaling
behavior of the characteristic temperatures,
\begin{equation}
\frac{1}{T(L)} = \frac{1}{T_{\rm c}} + \frac{D}{L^{1/\nu_1}} \;, \quad
L\rightarrow \infty \;,
\label{eq:T_L}
\end{equation}
where the coefficient~$D$ is nonuniversal and also depends on the
thermodynamic property for which $T(L)$ is determined. $\nu_1$ denotes
the critical exponent for the correlation length and takes the Ising
value~$0.630$.\cite{ising3d} Extrapolation of $T_Q(L)$ and $T_C(L)$
using Eq.~(\ref{eq:T_L}) yields virtually identical estimates for
$T_{\rm c}$.

\begin{figure}
\begin{center}
\includegraphics[width=\figurewidth]{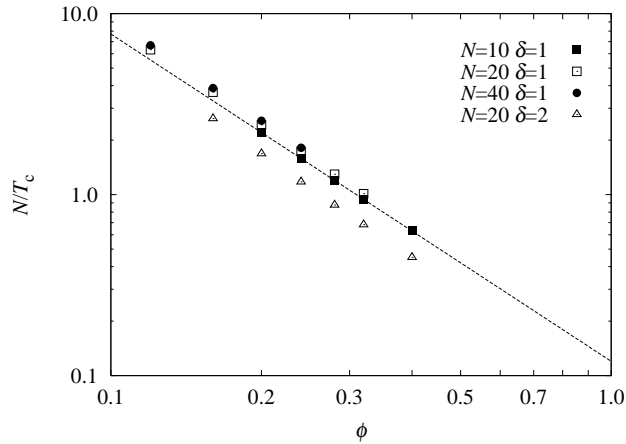}
\caption{\label{fig:Tc_phi_N}Scaling of the inverse critical demixing
temperature~$N/T_{\rm c}$ with concentration~$\phi$, plotted on a
double-logarithmic scale. The line, which represents a fit to the data
for $N=10$, describes a power-law dependence $T_{\rm c} \propto
\phi^{-1.808}$. The open triangles refer to data for a stronger
repulsion~$\delta$ between unlike monomers and hence lie at
systematically higher temperatures.}
\end{center}
\end{figure}

Having obtained the critical temperatures for the 15 cases listed in
Table~\ref{tab:systems}, we investigate their scaling with chain length
and concentration. As shown in Fig.~\ref{fig:Tc_phi_N}, for fixed chain
length and monomer repulsion~$\delta$, the inverse critical temperatures
exhibit a power-law dependence on~$\phi$,
\begin{equation}
\frac{N}{T_{\rm c}}\propto \frac{1}{\phi^{k}} \;.
\label{eq:my_Tc_phi}
\end{equation}
The results for $N=10$ can be described by an exponent $k=1.808~(5)$,
whereas least-square fits for $N=20$ and $N=40$ lead to $k=1.870~(3)$
and $k=1.880~(4)$, respectively. This clearly refutes the mean-field
result Eq.~(\ref{eq:FH_Tc}), according to which $T_{\rm c}$ increases
linearly with concentration ($k=1$). On the other hand, we note that our
result also shows a stronger variation with~$\phi$ than the nonlinear
relation Eq.~(\ref{eq:RG_Tc}). The blob-scaling arguments underlying the
theoretical prediction rely on the assumption that the radius of each
blob scales as $n_{\rm blob}^\nu$, where $n_{\rm blob}$ is the number of
monomers in a single blob. It would have been surprising if this scaling
behavior would be obeyed for the rather short chain lengths employed
here, which necessarily contain only few, relatively small blobs. In
particular, for small blobs the excluded-volume interactions are
partially screened, reducing the magnitude of the effective
exponent~$\nu_{\rm eff}$. Indeed, the observed values for
$k=1.80$--$1.88$ would correspond to an effective blob scaling
exponent~$\nu_{\rm eff}=0.55$--$0.56$, i.e., only a rather small
deviation from $\nu = 0.588$.  However, we note that our findings are
\emph{also} compatible with the alternative expression
Eq.~(\ref{eq:PRISM_Tc}) for an effective scaling exponent~$\nu_{\rm
eff}=0.54$--$0.57$.

Interestingly, Sariban \emph{et al.}~\cite{sariban94}, employing short
chains and a simple cubic lattice model, appear to have found excellent
agreement with a power law $k=1.60$, precisely matching the blob-scaling
result Eq.~(\ref{eq:RG_Tc}). Careful examination of Fig.~9 in
Ref.~\onlinecite{sariban94}, however, shows that all data that support
this scaling result were obtained for high monomer concentrations,
$\phi=0.40$, $0.60$, $0.80$ and chain lengths~$N \leq 32$.  For such
concentrated solutions, the blob scaling arguments certainly do not
apply, casting a doubt on the origin of the observed power-law behavior.
Inclusion of the data for $\phi=0.20$ leads to exponents that are
significantly larger in magnitude. We note that even for the same value
of the scaling variable employed in Ref.~\onlinecite{sariban94}
[$N\phi^{1/(3\nu-1)}$], the data for the longest chains ($N=64$) lead to
a power law that differs appreciably from $k=1.60$. This brings the
observations in Ref.~\onlinecite{sariban94} considerably closer to our
data for the bond fluctuation model. All data certainly appear
consistent with a slow approach of the predicted limiting behavior
Eq.~(\ref{eq:RG_Tc}), but it also cannot be excluded that
\emph{simultaneously} an alternative mechanism\cite{guenza97} applies
that leads to a rather comparable power-law behavior in the concentrated
solution regime.

Another aspect in Fig.~\ref{fig:Tc_phi_N} is the dependence of the
critical demixing temperature on the chain length.  Although the plotted
data for $\delta=1$, in which $T_{\rm c}$ is scaled by $N$, do not
exactly coincide, the remaining chain length dependence is weak and
suggestive of an additive small-$N$ correction that vanishes in the
limit $N\rightarrow \infty$.  In order to show the effect of the A--B
repulsion on $T_{\rm c}$, we also include data for $\delta=2$ for
$N=20$. These results exhibit the same power-law dependence on
concentration, but phase separation occurs at systematically higher
temperatures than for $\delta=1$, as expected from the stronger A--B
repulsion.

\subsection{Polymer--polymer phase separation and the role of the
dilute--semidilute threshold}
\label{sec:phi_star}

\begin{figure}
\begin{center}
\includegraphics[width=\figurewidth]{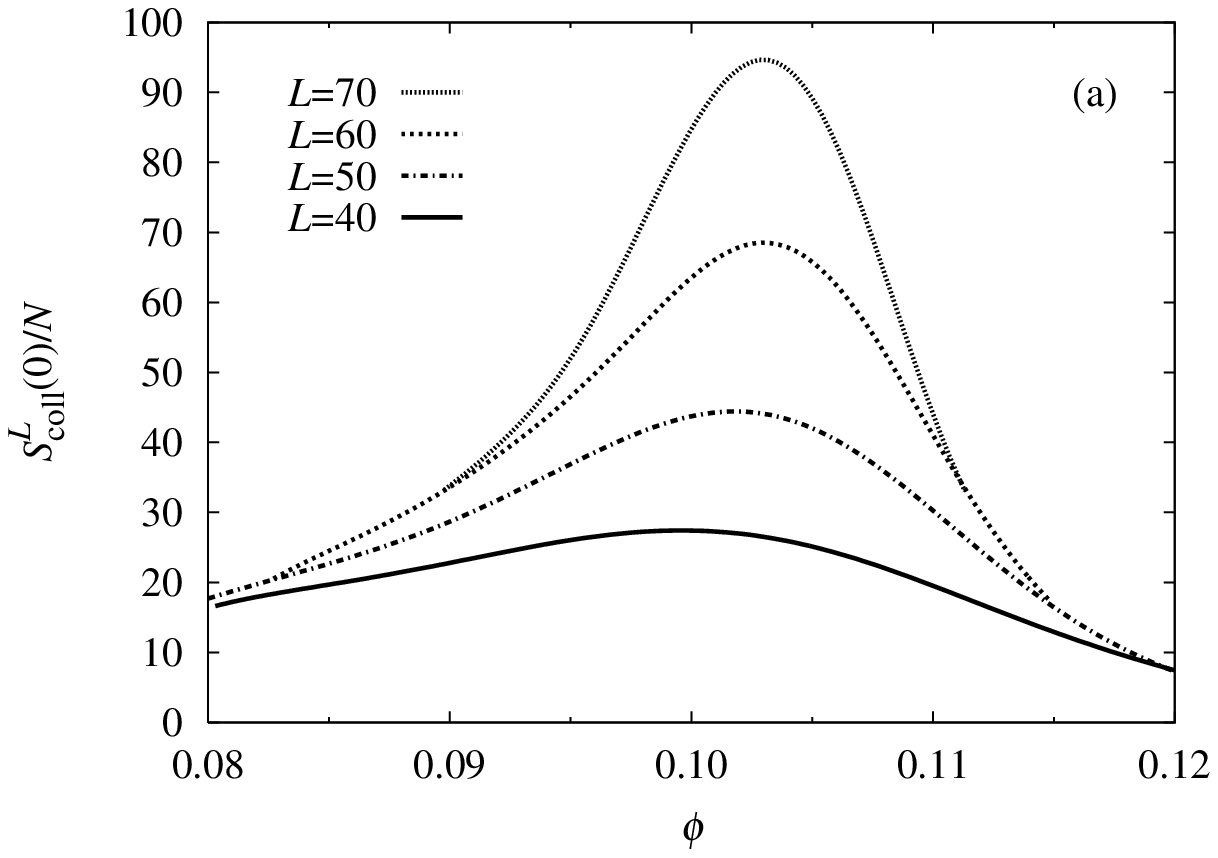}
\includegraphics[width=\figurewidth]{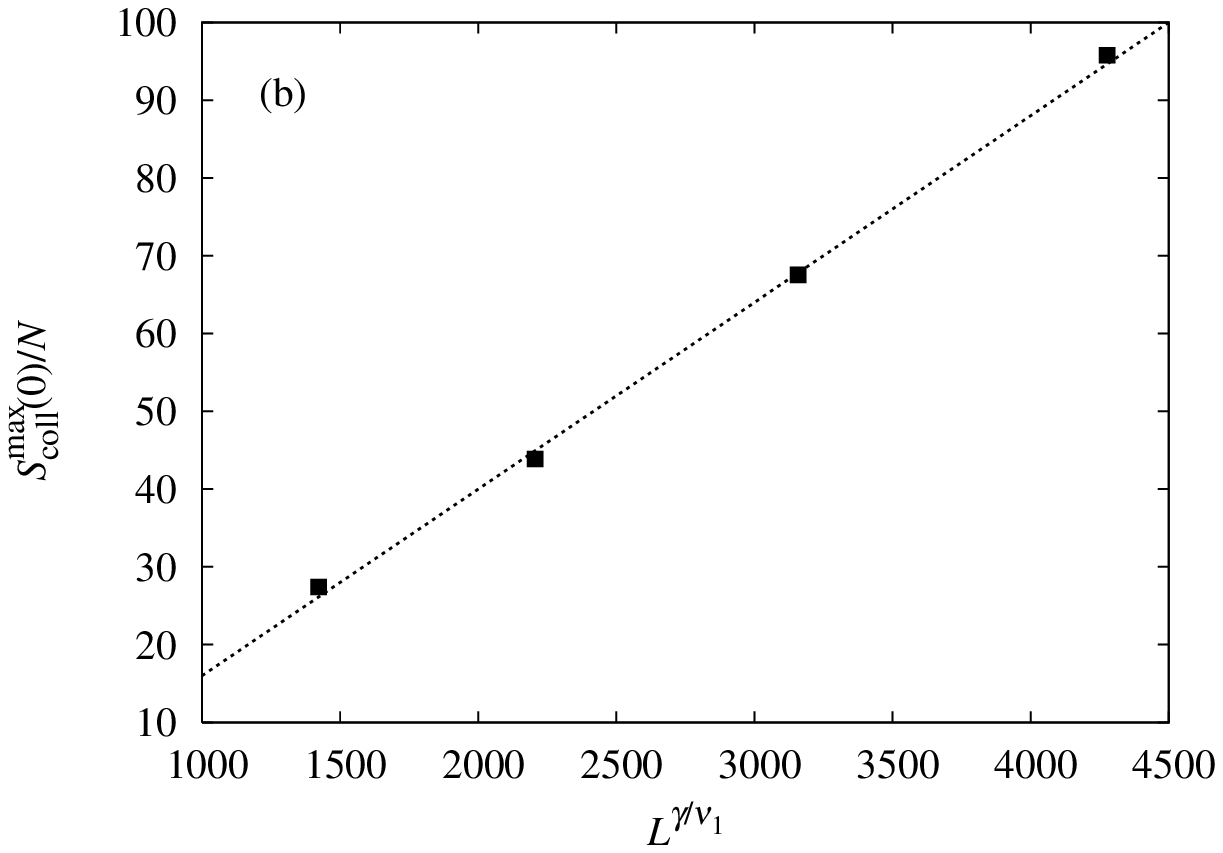}
\caption{\label{fig:chi_L}(a) Zero-angle scattering intensity~$S_{\rm
coll}(0)$ (scaled by the chain length~$N$) as a function of monomer
concentration~$\phi$ for a ternary solution containing two polymer
species with degree of polymerization~$N=20$. The data are obtained from
grand-canonical simulations at fixed (reduced) temperature~$T=0.05$.
Away from the maximum the data for different linear system sizes~$L$
converge rapidly, whereas the maximum itself exhibits strong finite-size
effects. (b) The height of the maximum increases with system size as
$L^{\gamma/\nu_1} \simeq L^{1.96}$, where $\gamma$ and $\nu_1$ are
critical exponents belonging to the Ising universality
class.\cite{ising3d} This confirms the occurrence of a critical phase
transition as a function of~$\phi$.  As discussed in the text,
polymer--solvent separation is suppressed by eliminating attractive
interactions between monomers of the same species, $\varepsilon_{\rm
AA}=\varepsilon_{\rm BB}=0$.}
\end{center}
\end{figure}

According to FH theory~\cite{scott49,flory53}, PP phase separation occurs at
arbitrarily low concentration. On the other hand, de
Gennes~\cite{degennes78,degennes79} has predicted that dilute solutions
($\phi<\phi^*$) exhibit no phase separation, whereas a critical
demixing transition takes place for symmetric systems if the A--B
repulsion is strong enough (or temperature sufficiently low).

\begin{figure}
\begin{center}
\includegraphics[width=\figurewidth]{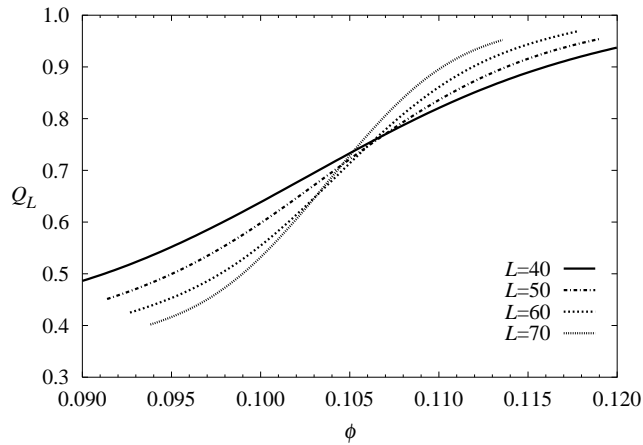}
\caption{\label{fig:Q_phi}Fourth-order amplitude ratio~$Q_L$ as a
function of monomer concentration for different system sizes~$L$. The
data pertain to the system described in the caption of
Fig.~\ref{fig:chi_L}. The crossing points, which occur near the location
of the compressibility maxima in Fig.~\ref{fig:chi_L}a, approach the
concentration at which critical polymer--polymer demixing takes place.}
\end{center}
\end{figure}

In order to test this scenario, we consider a ternary solution with
$N_{\rm A}=N_{\rm B}=N=20$ at very low, fixed temperature $T=0.05$
(strong repulsion regime). We perform simulations in the grand-canonical
ensemble at several chemical potentials that correspond to a range in
monomer concentration and we specifically monitor the zero-angle
scattering intensity $S_{\rm coll}(0)$, Eq.~(\ref{eq:scoll}), and the
fourth-order amplitude ratio~$Q_L$, Eq.~(\ref{eq:Q_L}). Since the
simulations are carried out for four different system sizes, $L=40$,
$50$, $60$, $70$, a critical transition will manifest itself via
finite-size effects. The compressibility will exhibit a maximum that
increases with system size according to a well-defined power law and the
amplitude ratio will exhibit a universal crossing point.  The
compressibility data shown in Fig.~\ref{fig:chi_L}a, which were obtained
by means of histogram reweighting, exhibit precisely this behavior. For
each system size, $S_{\rm coll}(0)$ displays a maximum $S_{\rm
coll}^{\rm max}(0)$ as a function of monomer concentration. This maximum
scales with~$L$ as $L^{\gamma /\nu_1} \simeq L^{1.96}$
(Fig.~\ref{fig:chi_L}b). The power law is characteristic for the
compressibility at a critical phase transition with a one-component
order parameter. We note that these simulations are performed at very
low temperatures. Nevertheless, phase separation only occurs when a
certain critical concentration has been reached. The fourth-order
amplitude ratio of moments of the order-parameter distribution confirms
these observations. As shown in Fig.~\ref{fig:Q_phi}, the curves for
$Q_L$ for different system sizes exhibit a crossing point at a
concentration that is close to the concentration of maximum
compressibility (Fig.~\ref{fig:chi_L}a). The crossing curves are very
similar to those commonly employed to determine critical temperatures;
the curves for successive system sizes do not all cross at a single
concentration owing to finite-size corrections.  However, it is
noteworthy that the value of $Q_L$ at the crossing point approaches the
universal Ising value $Q_{L} = 0.6233$ (cf.\ Ref.~\onlinecite{ising3d})
with increasing $L$.

\begin{figure}
\begin{center}
\includegraphics[width=\figurewidth]{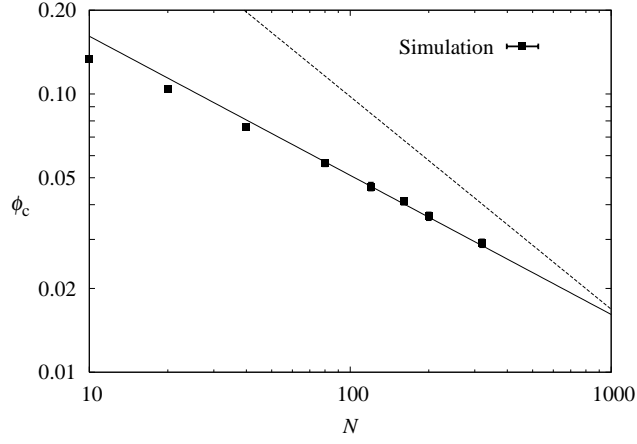}
\caption{Scaling of the critical concentration $\phi_{\rm c}$ with $N$
on a log--log scale. All results pertain to a system with
$\varepsilon_{\rm AA}=\varepsilon_{\rm BB}=0$ in the strong segregation
regime ($1/T=20$). The estimates for $N > 40$ were obtained from the
detection of crossing curves for $Q_L(T)$, as explained in the text
(cf.\ $\tilde{\phi}$ in Table~\ref{tab:phi_c}). The dashed line
represents the theoretical prediction that the critical concentration in
the strong segregation regime is proportional to the overlap threshold.
The solid line represents the effective power-law behavior observed for
the longest chains investigated in this work ($120 \leq N \leq 320$).}
\label{fig:phistar}
\end{center}
\end{figure}

Table~\ref{tab:phi_c} lists the critical concentration~$\phi_{\rm c}$
determined from both the compressibility and the fourth-order amplitude
ratio for chain lengths $10$, $20$, and~$40$. As an alternative
approach, we exploit the notion that, at fixed concentration~$\phi$, the
curves for $Q_L(T)$ for two different system sizes~$L$ must exhibit a
crossing point, if the system undergoes phase separation. Thus, if such
a crossing point is observed, $\phi \ge \phi_{\rm c}$ and if no crossing
point can be detected, $\phi < \phi_{\rm c}$. The resulting estimates
($\tilde{\phi}$ in Table~\ref{tab:phi_c}) indeed agree reasonably well
with those for~$\phi_{\rm c}$. Since the second method is
computationally more efficient, we have used it to estimate the critical
concentration for chain lengths as large as $N=320$, see
Fig.~\ref{fig:phistar}.

\begin{table}
\caption{Critical concentration for phase separation at low temperatures
(strong segregation factor). $\phi_{\rm c}$ is determined from
extrapolation of the maxima of the compressibility and of the
crossing points of the fourth-order amplitude
ratio~$Q_L$ (cf.\ Figs.\ \ref{fig:chi_L} and~\ref{fig:Q_phi}).
$\tilde{\phi}$ is determined by detecting whether, at fixed
concentration, two curves for $Q_L$ exhibit a crossing point at some
temperature.}
\label{tab:phi_c}
\begin{ruledtabular}
\begin{tabular}{ccc}
 $N$  & $\phi_{\rm c}$    & $\tilde{\phi}$        \\ \hline
 $10$ & $0.131 \pm 0.005$ & $0.13375 \pm 0.00125$ \\
 $20$ & $0.101 \pm 0.005$ & $0.10375 \pm 0.00125$ \\ 
 $40$ & $0.077 \pm 0.003$ & $0.07625 \pm 0.00125$ \\
\end{tabular}
\end{ruledtabular}
\end{table}

According to de Gennes~\cite{degennes79}, in a good solvent the overlap
threshold scales as
\begin{equation}
\phi^* \propto N/R_g^3 \propto N^{1-3\nu}=N^{-0.764} \;,
\label{eq:phistar_N}
\end{equation}
and for the low-temperature regime $\phi_{\rm c}$ is predicted to follow
the same scaling behavior. Our findings for $\phi_{\rm c}$
(Fig.~\ref{fig:phistar}) exhibit a slowly varying, effective power-law
behavior, which for the longest chains approaches $\phi_{\rm c} \propto
N^{-0.50}$. It is well possible that the trend in the effective exponent
continues and reaches the behavior predicted in
Eq.~(\ref{eq:phistar_N}). We note that the analysis of Broseta \emph{et
al.}\cite{broseta87}, which modifies the scaling of $\phi_{\rm c}$ from
Eq.~(\ref{eq:critconc}) to Eq.~(\ref{eq:critconc-rg}), only appears to
apply to the semidilute regime. This would imply different scaling
behavior for the weak and the strong segregation regimes, and it is
therefore not entirely clear whether de Gennes' original prediction for
the low-temperature regime ($\phi_{\rm c} \simeq \phi^*$) indeed
remains unaltered.

\begin{figure}
\begin{center}
\includegraphics[width=\figurewidth]{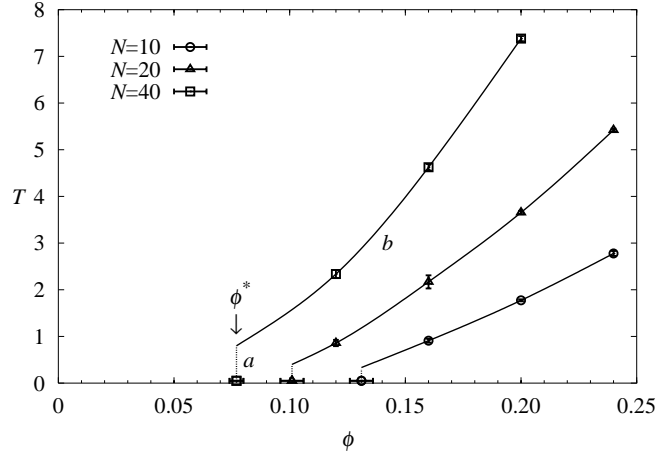}
\caption{Phase diagram for polymer--polymer separation as a function of
total monomer concentration and temperature, for different $N$. All data
pertain to systems without monomer attractions, $\varepsilon_{\rm
AA}=\varepsilon_{\rm BB}=0$, and hence differ from those presented in
Fig.~\ref{fig:Tc_phi_N}. The curves (which are drawn as guides to the
eye) separate the mixing region (left-hand side) from the demixing
region (right-hand side). The curve~$b$ indicates the weak segregation
regime, in which phase separation occurs at high concentrations, whereas
the dotted line~$a$ reflects the strong segregation regime where phase
separation takes place for concentrations above the overlap threshold.
(For clarity, only the curves for $N=40$ are labeled.)}
\label{fig:phi_star_Tc}
\end{center}
\end{figure}

Finally, the critical concentrations for the low-temperature regime can
be combined with the results for the systems listed in
Table~\ref{tab:systems} to assemble a phase diagram showing the critical
lines for three different chain lengths in the
concentration--temperature plane, see Fig.~\ref{fig:phi_star_Tc}. The
critical lines in the semidilute regime are terminated by the lines
labeled~$\phi^*$. These cutoff lines are drawn vertically to reflect the
observation that in this model $\phi^*$ is insensitive to temperature
variation.

\subsection{Critical amplitudes in ternary solutions}
\label{sec:amplitudes}

\begin{figure}
\begin{center}
\includegraphics[width=\figurewidth]{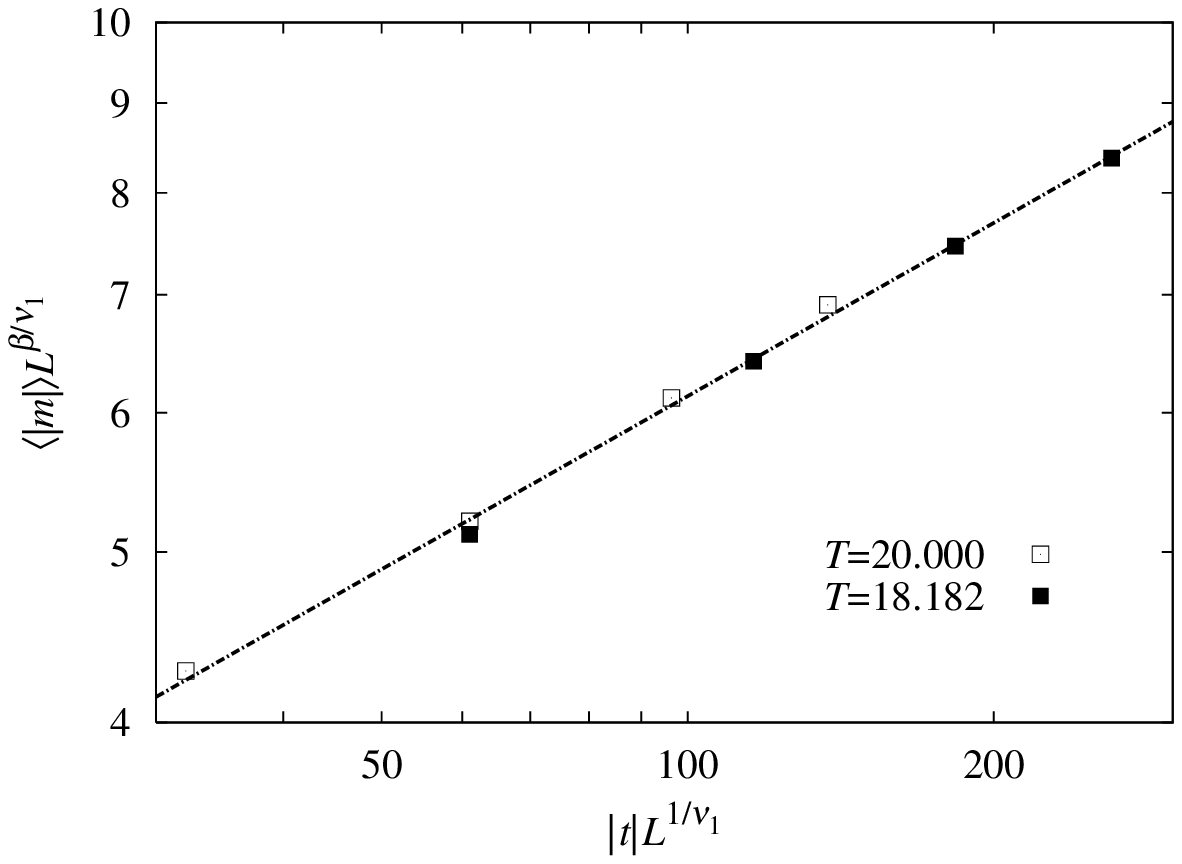}
\caption{\label{fig:mL}Finite-size scaling of the demixing order
parameter~$\langle |m|\rangle$ for a solution with $\phi=0.24$ and
$N=40$, near its critical temperature $T_{\rm c}=22.0~(1)$.  The two
data sets pertain to temperatures $T=18.182$ and $T=20.000$,
respectively. For each temperature, the data points correspond to system
sizes $L=40$, $60$, $80$, and $100$. All points are described by a
single power law~$(|t|L^{1/\nu_1})^\beta$, with $\beta = 0.327$.}
\end{center}
\end{figure}

\begin{figure}
\begin{center}
\includegraphics[width=\figurewidth]{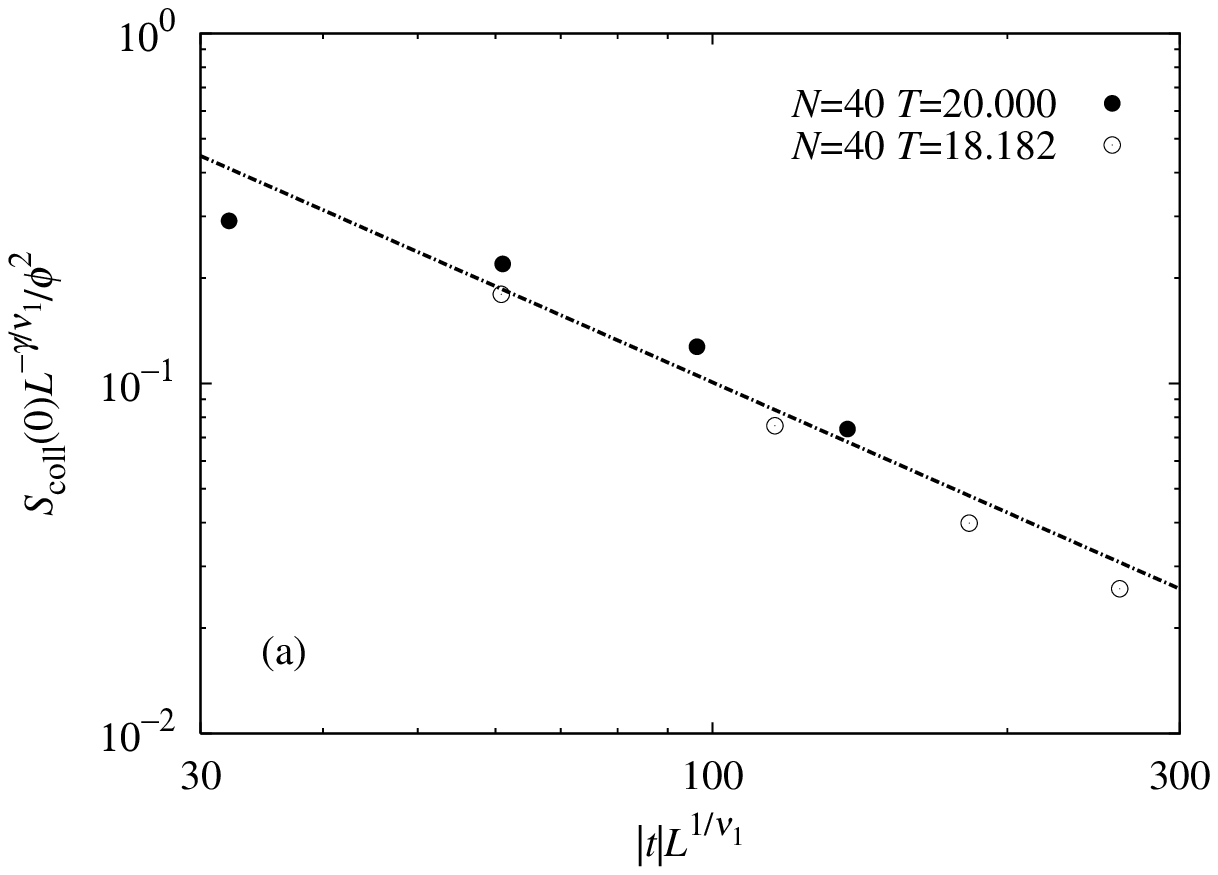}
\includegraphics[width=\figurewidth]{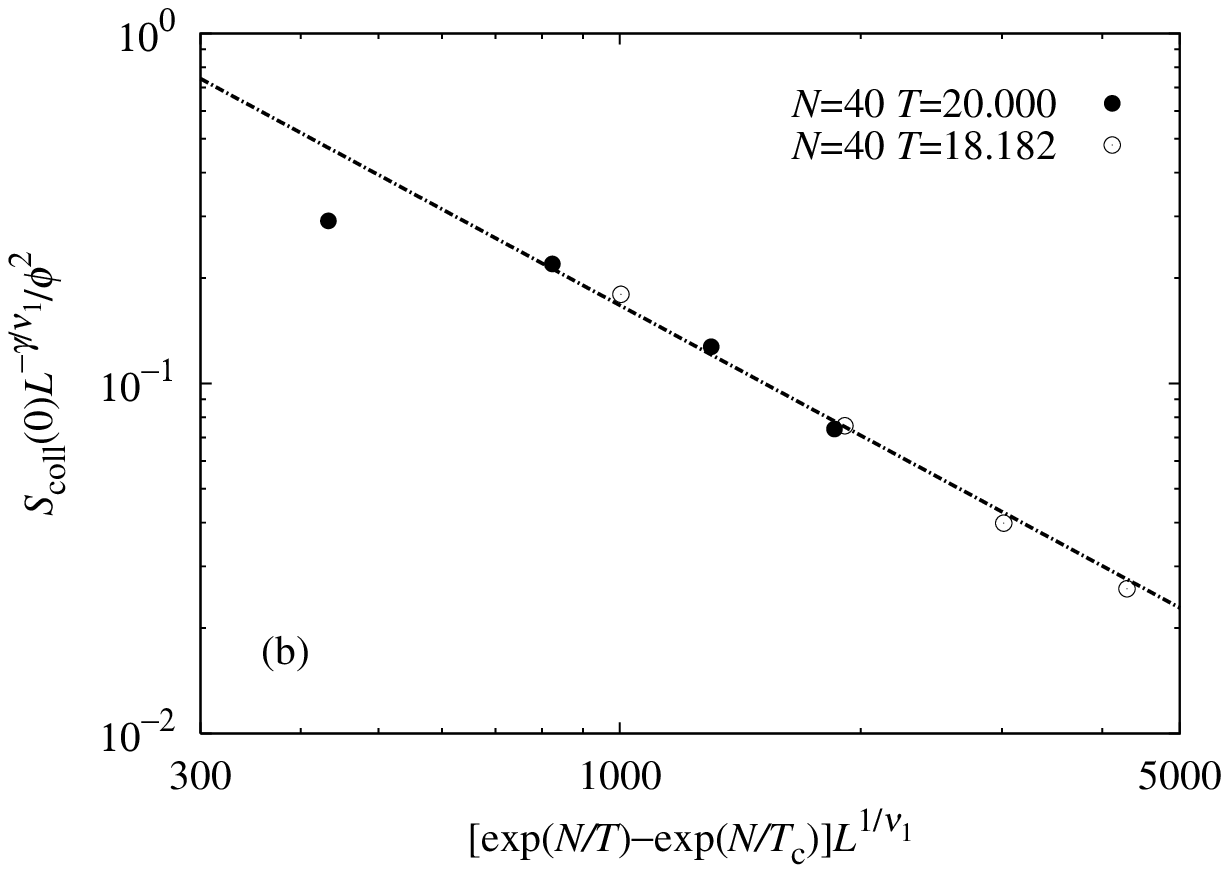}
\caption{\label{fig:SL}Finite-size scaling behavior of the zero-angle
scattering intensity $S_{\rm coll}(0)/\phi^2$ for the system described
in the caption of Fig.~\ref{fig:mL}. Panel~(a) employs the regular
reduced temperature $|(T-T_{\rm c})/T_{\rm c}|$, which does not permit a
description of all data by a single power law, presumably because of
corrections to scaling. In panel~(b), the alternative temperature
variable $[\exp(N/T)-\exp(N/T_{\rm c})]$ is used, which indeed greatly
improves the data collapse and permits the description of all data by a
single power law~$(|t|L^{1/\nu_1})^{-\gamma}$, with $\gamma = 1.237$.
The left-most data point in panel~(b) deviates because of finite-size
corrections.}
\end{center}
\end{figure}

Near a critical point, finite-size scaling theory implies
\begin{equation}
\langle |m|\rangle L^{\beta/\nu_1}=\tilde{f}(tL^{1/\nu_1})
\label{eq:m_L}
\end{equation}
and
\begin{equation}
S_{\rm coll}(0)L^{-\gamma/\nu_1}/\phi^2=\tilde{S}(tL^{1/\nu_1}) \;,
\label{eq:SL}
\end{equation}
where $\tilde{f}$ and $\tilde{S}$ are universal scaling
functions.\cite{note-phi2} In the finite-size scaling limit,
$tL^{1/\nu_1} \ll 1$, these equations yield the finite-size scaling
behavior at the critical point, commonly employed for the numerical
determination of critical exponents.\cite{ising3d} Outside the
finite-size scaling regime, $tL^{1/\nu_1} \gg 1$, but still sufficiently
close to the critical temperature to be within the scaling regime $t \ll
1$, the scaling functions reproduce the critical behavior in the
thermodynamic limit, i.e., $\tilde{f}(x) \sim x^\beta$ and $\tilde{S}(x)
\sim x^{-\gamma}$. Here, we focus on this second regime.  Assuming Ising
values for the exponents, we plot $\langle |m|\rangle L^{\beta /\nu_1}$
in Fig.~\ref{fig:mL} as a function of $|t|L^{1/\nu_1}$ for a solution
with $N=40$ and total concentration~$\phi=0.24$.  For this system, which
has a critical temperature $T_{\rm c}=22.0~(1)$, two sets of data are
plotted, obtained at $T=18.182$ and $T=20.000$, respectively. Each set
contains four different system sizes~$L$, and all data turn out to
collapse on a single line which, on this double-logarithmic scale, has a
slope~$0.327$, the critical exponent~$\beta$ in the Ising universality
class.  This confirms Eq.~(\ref{eq:m_L}) and our assumption of Ising
exponents.  Likewise, data for the zero-angle scattering
intensity~$S_{\rm coll}/\phi^2$ are plotted in Fig.~\ref{fig:SL}a. The
data do not fall onto a single curve as well as the data for the order
parameter do, which can possibly be ascribed to the fact that the
reduced temperature for our data is relatively large and hence
corrections to scaling start to become important. Therefore, we replot
the same data in Fig.~\ref{fig:SL}b as a function of an alternative
temperature variable $[\exp(N/T)-\exp(N/T_{\rm c})]$, proposed in
Ref.~\onlinecite{sariban87}. To leading order, this variable equals
$-[(N/T_{\rm c}) \exp(N/T_{\rm c})]t$. Note that, as implied by
Eq.~(\ref{eq:my_Tc_phi}), the prefactor is independent of~$N$.  This
variable indeed improves the scaling behavior and the data in
Fig.~\ref{fig:SL}b are well described by a power law
$(tL^{1/\nu_1})^{-\gamma}$, with $\gamma = 1.237$. Similar scaling
analyses have been carried for our results for $N=10$ and~$N=20$ (not
shown).

\begin{figure}
\begin{center}
\includegraphics[width=\figurewidth]{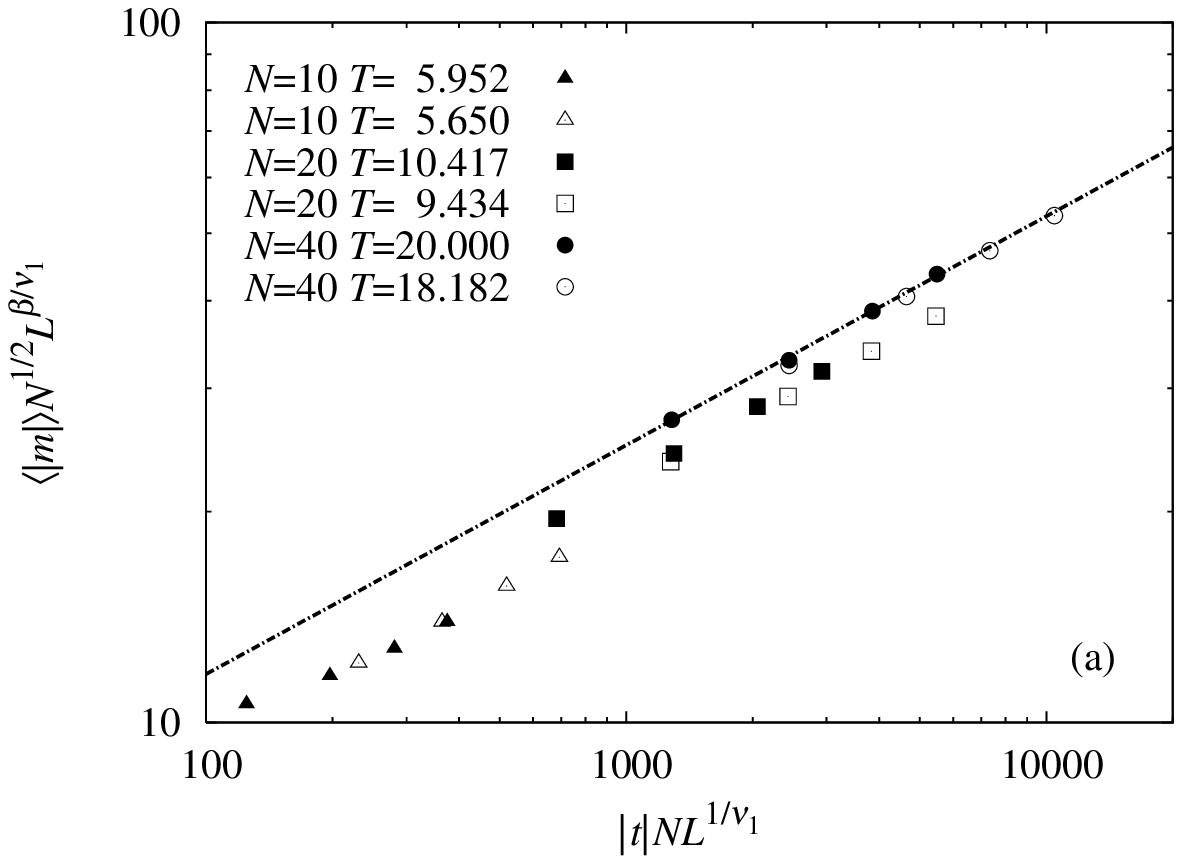}
\includegraphics[width=\figurewidth]{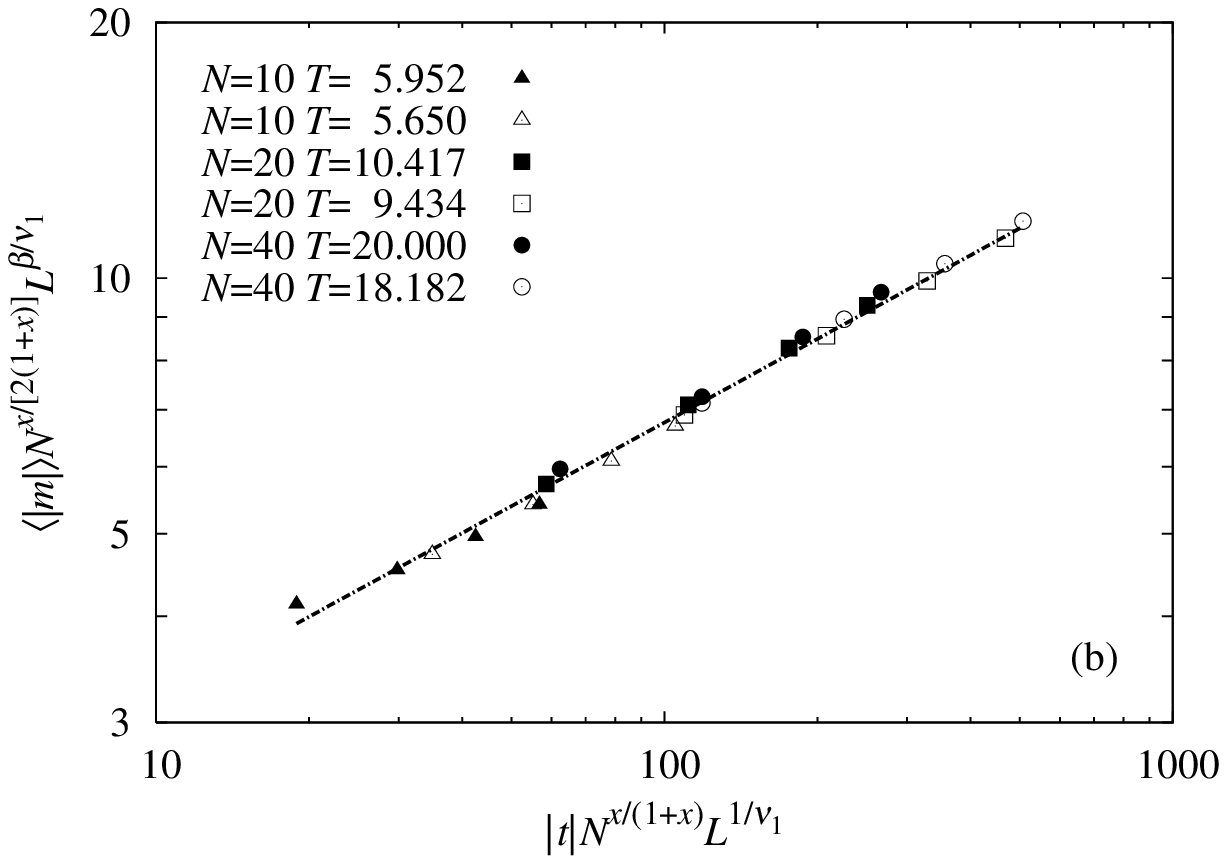}
\caption{\label{fig:mN}Scaling of the near-critical order parameter
$\langle |m|\rangle$ with system size~$L$ and degree of
polymerization~$N$.  Panel~(a) demonstrates that the conventional
scaling [Eq.~(\ref{eq:mN})] is not applicable in the semidilute regime.
In panel~(b), a modified scaling of the critical amplitude is adopted
[Eq.~(\ref{eq:mN_RG})], which leads to an excellent collapse of all data
points onto a single power law with Ising exponent~$\beta=0.327$.  Data
points pertain to the following systems (all at concentration
$\phi=0.24$): (i)~$N=10$ ($T_{\rm c}=6.309$): $T=5.952$ and $T=5.650$.
For each temperature, system sizes $L=30$, $40$, $50$, and $60$ are
employed.  (ii)~$N=20$ ($T_{\rm c}=11.547$): $T=10.417$ and $T=9.434$.
For each temperature, system sizes $L=40$, $60$, $80$, and $100$ are
employed.  (iii)~$N=40$ ($T_{\rm c}=22.0$): $T=20.000$ and $T=18.182$.
For each temperature, system sizes $L=40$, $60$, $80$, and $100$ are
employed.}
\end{center}
\end{figure}

By combining the data for different chain lengths, we can test the
modified scaling of $\hat{B}(N)$ and $\hat{\Gamma}(N)$ with $N$ as
proposed by Broseta, i.e., Eqs.\ (\ref{eq:mN_RG}) and~(\ref{eq:SN_RG}).
Figure~\ref{fig:mN} constitutes the counterpart of Fig.~\ref{fig:mL},
displaying the scaling of the order parameter~$\langle |m|\rangle$
outside the finite-size scaling regime (but within the critical region).
In Fig.~\ref{fig:mN}a, the conventional scaling with~$N$ is adopted, in
which the reduced temperature~$t$ is replaced by $tN \propto t/G$ and $\langle
|m|\rangle$ is multiplied by $N^{1/2}$, following Eq.~(\ref{eq:mN}).
Evidently, this does not properly describe the scaling of the critical
amplitude with the degree of polymerization.  However, replacing $N$ by
$N^{x/(1+x)}$, i.e., plotting $\langle |m|\rangle
N^{x/[2(1+x)]}L^{\beta/\nu_1}$ as a function of
$|t|N^{x/(1+x)}L^{1/\nu_1}$ (Fig.~\ref{fig:mN}b) leads to an excellent
collapse of all data points onto a single line that describes a power
law with exponent~$\beta = 0.327$. Thus, the order parameter scales as
\begin{equation}
\langle |m|\rangle N^{x/[2(1+x)]}L^{\beta/\nu_1} \propto
\left[ |t| N^{x/(1+x)}L^{1/\nu_1} \right]^{\beta} \;,
\end{equation}
which can be simplified to
\begin{equation}
\langle |m|\rangle \propto N^{x(\beta-0.5)/(1+x)}t^{\beta} \;,
\end{equation}
in agreement with Eq.~(\ref{eq:mN_RG}).

\begin{figure}
\begin{center}
\includegraphics[width=\figurewidth]{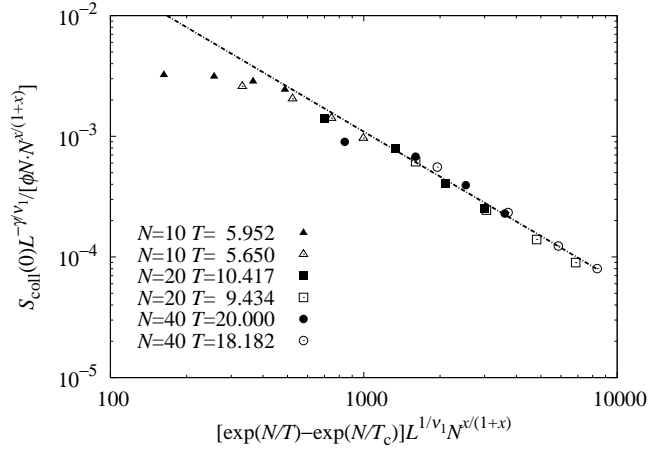}
\caption{\label{fig:SN_RG}Scaling of the zero-angle scattering intensity
near criticality. Data are assembled for all systems listed in the
caption of Fig.~\ref{fig:mN}. Except for some data points that cross
over into the finite-size scaling regime (left-hand side of the plot),
all data are described by a single power law with exponent~$-1.237$.
This confirms the modified scaling with degree of polymerization, see
Eq.~(\ref{eq:SN_RG}).}
\end{center}
\end{figure}

In a similar fashion we test Eq.~(\ref{eq:SN_RG}). Based upon
Fig.~\ref{fig:SL} we use the modified reduced temperature
$[\exp(N/T)-\exp(N/T_{\rm c})]$ instead of $-t$, which does not affect
the scaling with~$N$. Since all data points apply to the same concentration,
we investigate the scaling of $S_{\rm coll}(0)/(\phi
N)$. Figure~\ref{fig:SN_RG} demonstrates that all data points except for
those with the smallest values of $|t| L^{1/\nu_1}$ (which cross over to
a horizontal line representing the critical finite-size amplitude) are
described by a power law with exponent $-\gamma=-1.237$. Thus, the
following scaling behavior is recovered,
\begin{equation}
\frac{S_{\rm coll}(0)}{\phi N} L^{-\gamma/\nu_1} N^{-x/(1+x)}
\propto \left[ |t| L^{1/\nu_1} N^{x/(1+x)} \right]^{-\gamma}
\end{equation}
which can be simplified to
\begin{equation}
\frac{S_{\rm coll}(0)}{\phi N} \propto N^{x(1-\gamma)/(1+x)} t^{-\gamma} \;,
\end{equation}
consistent with Eq.~(\ref{eq:SN_RG}).

Thus, we conclude that the chain-length dependence of the critical
amplitudes in the semidilute regime can \emph{not} be described by the
conventional scaling laws, but that the modified scaling behavior
proposed by Broseta \emph{et al.}\cite{broseta87} provides an excellent
description already for relatively short chains.

\subsection{Shape variation of polymers upon polymer--polymer phase
separation}
\label{sec:shape}

In our earlier communication\cite{guo03a}, we studied the shape change
of polymers upon phase separation in terms of the \emph{asphericity}~$A$
(see Eq.~(\ref{eq:A})). We confined ourselves to isothermal variation of
the concentration in the strong segregation regime, i.e., phase
separation near $\phi = \phi^*$. This corresponds to an isotherm that
intersects the line~$a$ in Fig.~\ref{fig:phi_star_Tc}. Along such an
isotherm, the asphericity of the majority component in a given phase
decreases merely slightly at sufficiently high concentration due to the
screening of the excluded-volume interactions (thus, the shape of a coil
becomes slightly more spherical upon increase of the concentration).  By
contrast, the minority component in each phase exhibits a strong
variation of the asphericity. Since a typical coil belonging to the
minority species is surrounded by polymers of the majority species, the
strong repulsion ``squeezes'' the minority coil to a much more spherical
shape, as confirmed by a rapid drop in $A$ for the minority polymers.
Here, we extend this work by studying other regions of the phase diagram,
namely (i) variation of the total monomer concentration at higher
temperatures (such that phase separation takes place in the semidilute
regime) and (ii) temperature variation at fixed concentration (for $\phi
> \phi^*$), where we can also verify any influence of attractions
between monomers of the same species (in Ref.~\onlinecite{guo03a}, such
interactions were explicitly ruled out, in order to prevent
polymer--solvent phase separation near the overlap concentration).  In
addition, we extend our analysis of coil shapes by means of the
parameter~$S$ [Eq.~(\ref{eq:S})], which distinguishes between
prolate-ellipsoidal and oblate-ellipsoidal shapes.

\begin{figure}
\begin{center}
\includegraphics[width=\figurewidth]{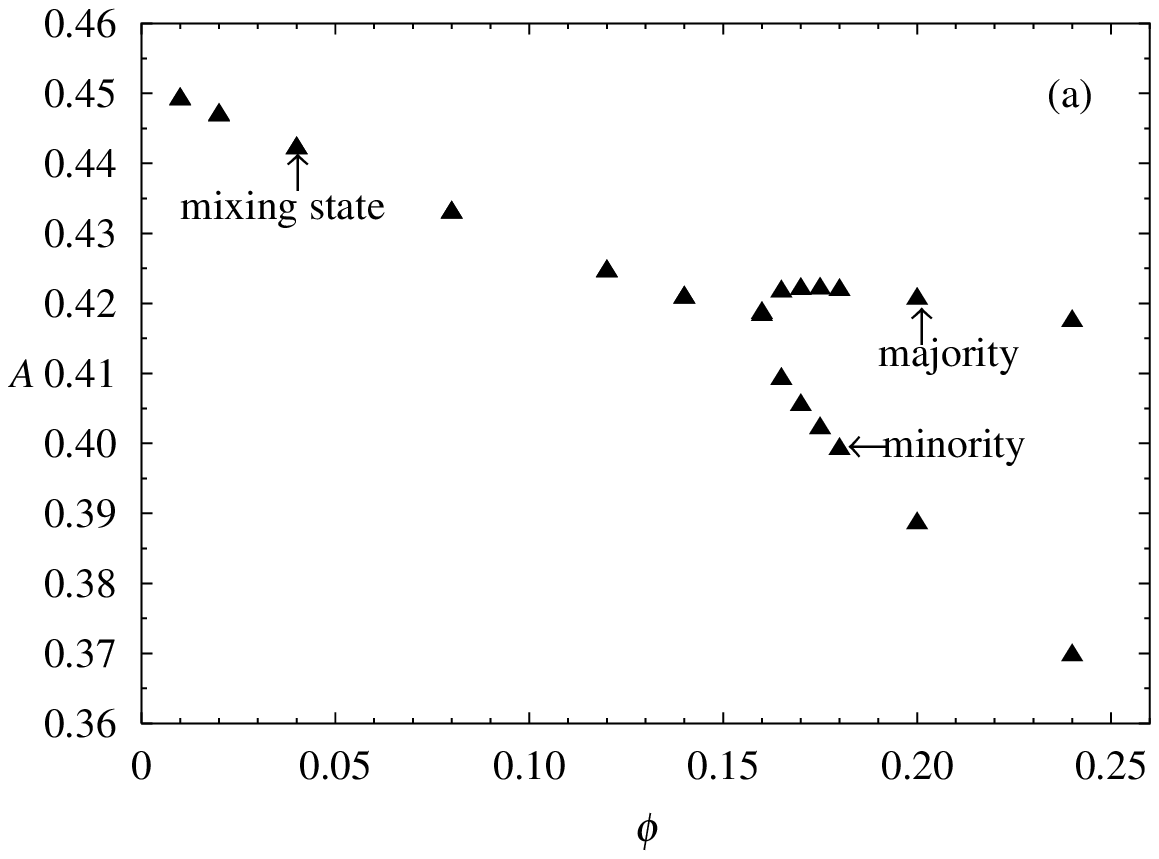}
\includegraphics[width=\figurewidth]{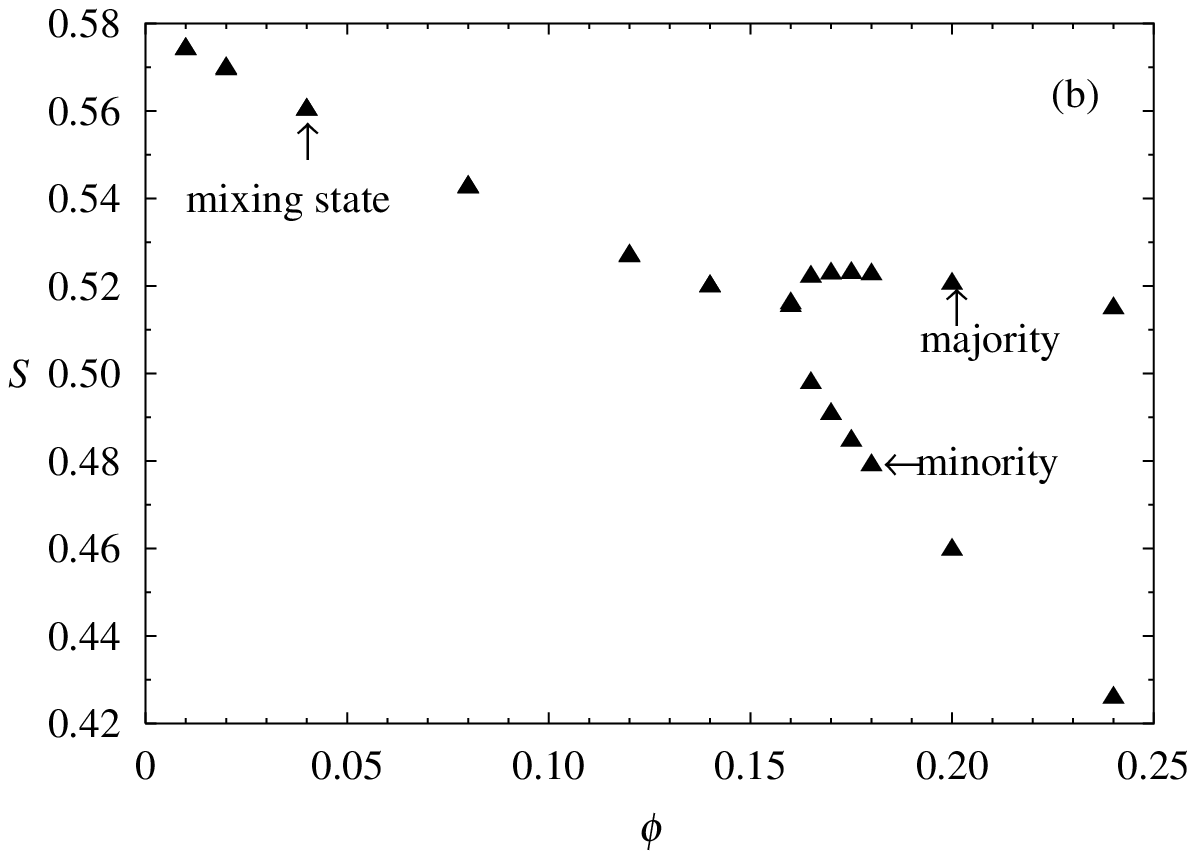}
\caption{Characterization of the coil shape in a ternary solution, as a
function of monomer concentration~$\phi$. The temperature is fixed at
$T=2.14$ and the results pertain to a symmetric mixture with chain
length~$N=20$. Phase separation occurs near~$\phi=0.16$ and is driven
purely by repulsions between unlike monomers ($\varepsilon_{\rm
AA}=\varepsilon_{\rm BB}=0$). (a) The asphericity~$A$ [Eq.~(\ref{eq:A})]
decreases with increasing concentration, in particular for the minority
component, indicating that the coils take an increasingly spherical
shape. (b) The oblate--prolate parameter~$S$ [Eq.~(\ref{eq:S})]
demonstrates that the coils have a prolate shape over the entire
concentration range. The ``elongated'' shape is most pronounced in the
dilute regime.}
\label{fig:N20_highT}
\end{center}
\end{figure}

\begin{figure}
\begin{center}
\includegraphics[width=\figurewidth]{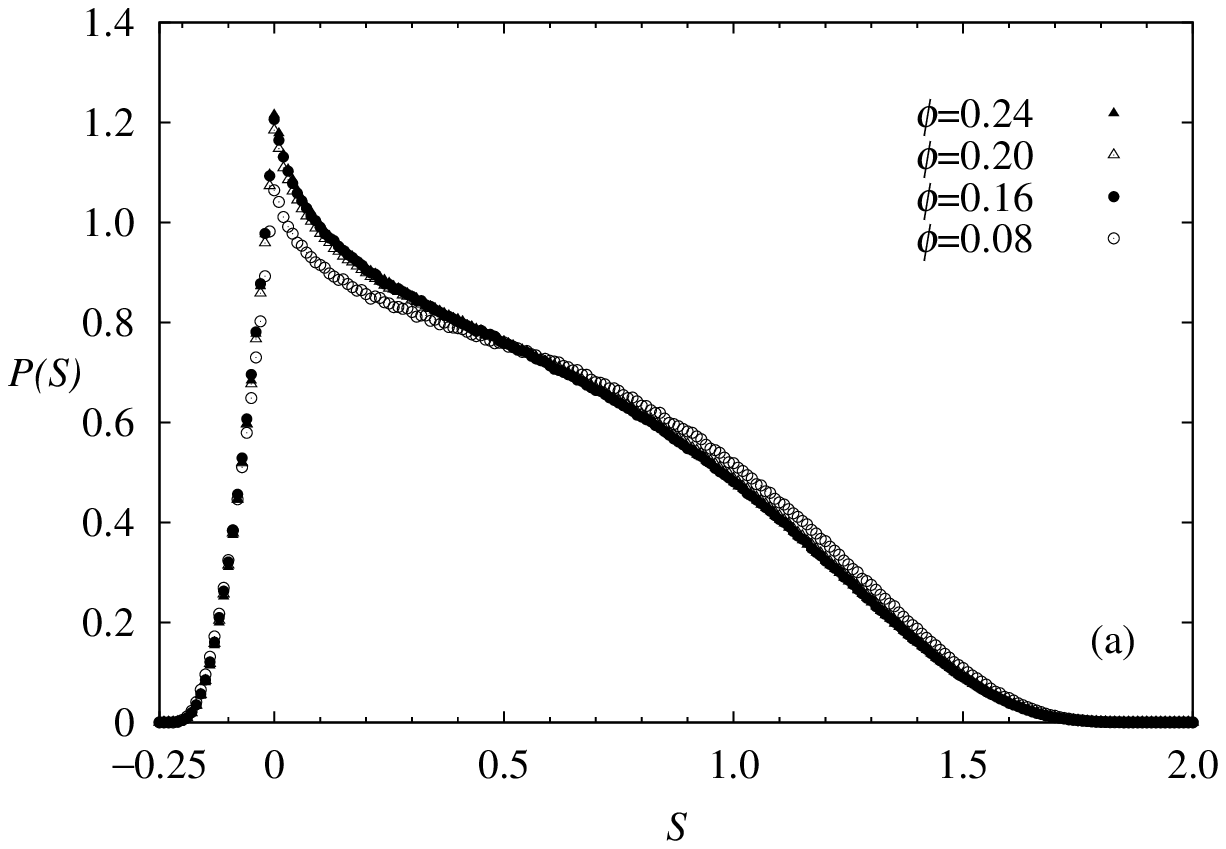}
\includegraphics[width=\figurewidth]{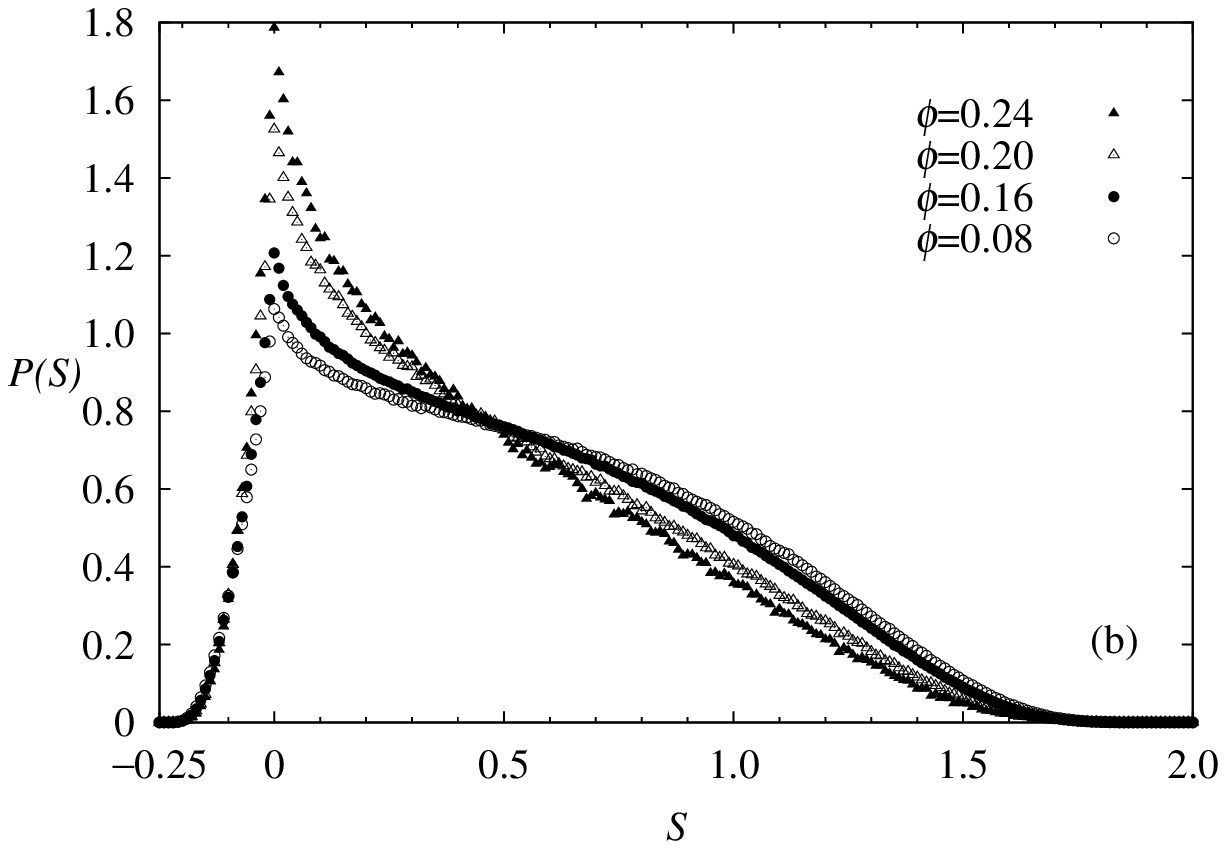}
\caption{\label{fig:N20_S3_dist}Probability distribution of the
oblate--prolate parameter~$S$ of (a) the majority component and (b) the
minority component for different concentrations~$\phi$, at $T=2.14$ for
$N=20$. All data pertain to $\varepsilon_{\rm AA}=\varepsilon_{\rm
BB}=0$. While the majority component exhibits a virtually invariant
distribution for concentrations in the demixed regime $\phi \gtrsim
0.16$, the minority component shows a significant sharpening of the peak
at $S=0$, which corresponds to a spherical coil shape.}
\end{center}
\end{figure}

Figure~\ref{fig:N20_highT}a shows the variation of $A$ for a symmetric
mixture with $N=20$ along the $T=2.14$ isotherm, which crosses the
critical line in the semidilute regime (cf. Fig.~\ref{fig:phi_star_Tc}).
Just as observed for the strong segregation regime\cite{guo03a}, the
asphericity for the majority component decreases slowly while for the
minority component it drops rapidly. The location of the bifurcation
in~$A$ is in good agreement with the corresponding critical
concentration~$\phi_{\rm c}$ in Fig.~\ref{fig:phi_star_Tc}.  The nature
of the aspherical shape is characterized further by the variation of the
oblate--prolate parameter~$S$ over this concentration range
(Fig.~\ref{fig:N20_highT}b). As the concentration increases in the
dilute regime, the coil shape becomes less prolate, until the
concentration has reached its critical value and phase separation
occurs. Beyond $\phi_{\rm c}$, the minority component (e.g., a chain of
type~B in the A-rich phase) then becomes less prolate at an even higher
rate. On the other hand, the majority component becomes \emph{less
spherical} and \emph{more prolate} immediately after phase separation.
As explained in Ref.~\onlinecite{guo03a}, we ascribe this to the
diminished repulsion that a typical majority chain experiences in a
homogeneous (unmixed) phase. Although the concentration dependence of
$A$ and~$S$ is very similar, there are significant differences in the
probability distributions for these quantities. While the distributions
for $A$ are relatively broad\cite{guo03a}, the distributions for $S$,
both for the majority component (Fig.~\ref{fig:N20_S3_dist}a) and for
the minority component (Fig.~\ref{fig:N20_S3_dist}b) exhibit a sharp
peak at zero (spherical shape). For the majority component the
distribution, including the peak, exhibits only minor variation with
concentration in the dilute regime and remains virtually unchanged for
$\phi > \phi_{\rm c}$. However, for the minority component the most
significant changes in the distribution, including a significant
sharpening of the peak, occur for concentrations in the unmixed regime
$\phi \gtrsim 0.16$.  For both components, the distributions provide
insight into the relative occurrence of various coil shapes, which can
also be oblate-ellipsoidal ($S < 0$).

\begin{figure}
\begin{center}
\includegraphics[width=\figurewidth]{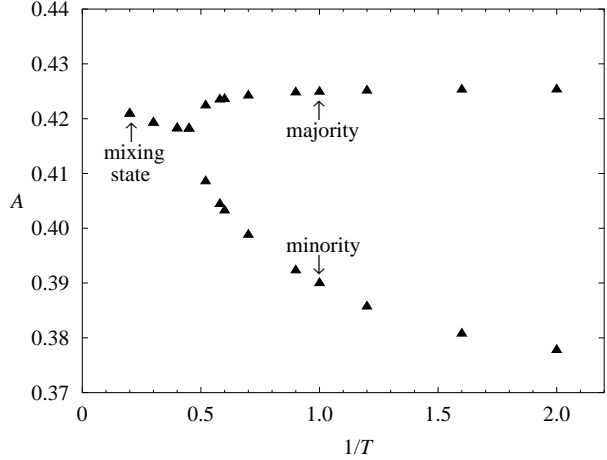}
\caption{\label{fig:N20_A3_phi016}Variation of the asphericity as a
function of inverse temperature for fixed concentration~$\phi=0.16$. The
results pertain to a system with $N=20$ and $\varepsilon_{\rm
AA}=\varepsilon_{\rm BB}=0$. Once phase separation is complete (i.e., at
low temperature) the majority component is essentially in an athermal
one-component solution.}
\end{center}
\end{figure}

As evident from the phase diagram (Fig.~\ref{fig:phi_star_Tc}), phase
separation can also be induced by lowering the temperature at fixed
concentration. The resulting variation in coil shape is illustrated in
Fig.~\ref{fig:N20_A3_phi016} for a system with $N=20$ and $\phi=0.16$. A
bifurcation similar to Fig.~\ref{fig:N20_highT}a is observed, at a
temperature that agrees with the critical temperature along the
corresponding isochore in Fig.~\ref{fig:phi_star_Tc}. Unlike the
behavior upon variation of the concentration, the asphericity of the
majority component now remains constant in the unmixed phase. Since the
concentration is constant, the screening of the excluded-volume
interactions remains unchanged, and the chains only become somewhat more
aspherical as phase separation continues, owing to the diminishing
repulsion from the minority component. Since all attractions between
monomers of the same species have been set to zero, the majority
component finds itself, once phase separation is complete, essentially
in an athermal one-component solution.

\begin{figure}
\begin{center}
\includegraphics[width=\figurewidth]{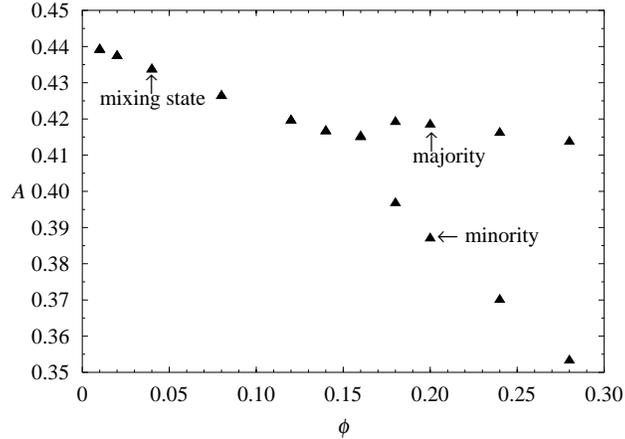}
\caption{\label{fig:N20_A3_thermal}Asphericity~$A$ as a function of
concentration for systems with an explicit, temperature-dependent
attraction between monomers of the same species, $\varepsilon_{\rm
AA}=\varepsilon_{\rm BB}=-1/T$. Chain length is set to $N=20$ and
temperature to $T=5.39$. This confirms that the absence of A--A and B--B
attractions does not qualitatively affect the results presented in
Figs.\ \ref{fig:N20_highT} and~\ref{fig:N20_S3_dist}.}
\end{center}
\end{figure}

The effect of attractions between monomers of the same species is
addressed in Fig.~\ref{fig:N20_A3_thermal}, for a system with
$\varepsilon_{\rm AA}=\varepsilon_{\rm BB}=-1/T$. The overall behavior
of the asphericity is very similar to that displayed in
Fig.~\ref{fig:N20_highT}a, with a small, systematic lowering of the
asphericity in the dilute regime, induced by the mutual attractions.
Thus, the results presented above (Figs.\ \ref{fig:N20_highT}
and~\ref{fig:N20_S3_dist}) and in Ref.~\onlinecite{guo03a} are not
qualitatively affected by the absence of monomer--monomer attractions.

\section{Conclusions}

We have studied polymer--polymer phase separation in a common
(nonselective) good solvent by means of Monte Carlo simulations of the
bond fluctuation model. Our calculations cover both dilute and
semidilute solutions, making it possible to distinguish critical
demixing in the weak and strong segregation regimes. In the weak
segregation regime we have determined the nonlinear concentration
dependence of the critical temperature by means of semi-grand-canonical
simulations and in the strong segregation regime, where phase separation
occurs upon variation of total monomer concentration rather than
temperature, by means of grand-canonical simulations. We observed a
sudden drop in critical temperature near the overlap threshold, as first
predicted by de Gennes\cite{degennes78}, although the chain-length
dependence of the corresponding critical concentration differed rather
strongly from that expected for the overlap concentration. Also the
nonlinear relation between critical temperature and critical
concentration in the semidilute regime exhibited a power law that
differs from the theoretical prediction\cite{schaefer85,broseta87},
which we ascribe to finite chain-length effects. However, it is also
possible that, for more concentrated solutions, similar nonlinear
behavior results from a concentration dependence of the local
correlation hole.\cite{guenza97} This may aid in explaining earlier
numerical results\cite{sariban94} and experiments on diblock copolymer
solutions\cite{lodge95b}, although there are also quantitative
differences between those results and the prediction of
Ref.~\onlinecite{guenza97}.  The modified chain-length dependence of all
critical amplitudes, first predicted by Broseta \emph{et
al.}\cite{broseta87} using a renormalization-group approach, has been
verified explicitly for the demixing order parameter and, for the first
time, for the zero-angle scattering intensity (osmotic compressibility),
whereas a scenario in which the critical amplitudes retain their
unmodified chain-length dependence can be convincingly ruled out.  The
observation of unrenormalized critical exponents is consistent with the
prediction that Fisher renormalization of those exponents only takes
place within a very narrow temperature range around the critical
temperature.  Conversely, the observation of nonclassical critical
exponents within a rather large temperature range is consistent with a
modified Ginzburg criterion\cite{broseta87}, which implies a slow
crossover to classical critical exponents.

In a preliminary report\cite{guo03a}, we observed that phase separation
causes polymer coils belonging to the minority component to become more
spherical, due to repulsion from the surrounding polymers of the
opposite species. Here, we have recovered this behavior for more general
monomer--monomer interactions and for simulations in which phase
separation occurs in the weak segregation regime, as well as for the
situation in which phase separation is induced by means of temperature
variation rather than variation of the total monomer concentration.
In addition, we have characterized the shape variation more precisely by
means of the distribution of the prolate--oblate parameter.

\begin{acknowledgments}
  This work is supported by the American Chemical Society Petroleum
  Research Fund under Grant No.\ 38543-G7 and by the National Science
  Foundation through an ITR grant (DMR-03-25939) via the Materials
  Computation Center at the University of Illinois at Urbana-Champaign.
  E.L. thanks Kurt Binder for drawing his attention to polymer--polymer
  separation in ternary solutions and Ken Schweizer for pointing out the
  connection with diblock copolymer solutions. The authors thank Intel
  for a generous equipment donation and M. M\"uller for providing parts
  of the simulation code. The recoil-growth code was originally
  developed with support of a postdoctoral Fellowship from the Max
  Planck Institute for Polymer Research and with support of the European
  Commission through TMR Grant No.\ ERB FMGE CT950051 for a TRACS visit
  to the Edinburgh Parallel Computing Centre.
\end{acknowledgments}


\begin{thebibliography}{60}
\bibitem[{not()}]{note-phi2}
\bibinfo{note}{The factor~$\phi^2$ was divided out on the basis of
  Eq.~(\protect\ref{eq:scoll}). Since we test the finite-size scaling behavior,
  Eq.~(\protect\ref{eq:SL}), for data at fixed concentration, this factor is
  immaterial in the present context.}
\end{thebibliography}


\end{document}